\RequirePackage{fix-cm}
\RequirePackage{amsmath}
\documentclass[smallextended]{svjour3}       % onecolumn (second format)
\smartqed  % flush right qed marks, e.g. at end of proof
\usepackage{graphicx}
\usepackage{amsfonts,amssymb}
\usepackage{enumerate}
\usepackage{tikz}
\usepackage{url}
\usepackage{bm} %    enables bold math symbols  e.g.  \bm{\gamma}
\usepackage[T1]{fontenc}
\usepackage{stmaryrd}
\usepackage{wrapfig}
\usepackage{hyperref}
\usepackage[title]{appendix}
\spnewtheorem{prop}[theorem]{Proposition}{\bfseries}{\itshape}

\spnewtheorem{defn}[theorem]{Definition}{\bfseries}{\itshape}

\spnewtheorem{remrk}[theorem]{Remark}{\bfseries}{\itshape}

\def\H{\mathbf{H}}

\def\x{\mathbf{x}}
\def\y{\mathbf{y}}
\def\z{\mathbf{z}}
\def\u{\mathbf{u}}

\def\s{\mathbf{s}}

\def\N{\mathcal{N}}

\def\E{\mathcal{E}}

\def \seff {\sigma_\mathrm{eff}^2}
\def \sxi{\sigma_\xi^2}
\def\CHI{\boldsymbol{\chi}}

\newcommand{\argmin}{\operatornamewithlimits{argmin}}
 \journalname{J. Stat. Phys.}
\begin{document}

\title{Critical Behavior and Universality Classes for an Algorithmic Phase Transition in Sparse Reconstruction%\thanks{Grants or other notes
}

\titlerunning{Critical Behavior in Sparse Reconstruction}        % if too long for running head

\author{Mohammad Ramezanali         \and
        Partha P. Mitra                              \and
         Anirvan M. Sengupta %etc.
}


\institute{M. Ramezanali \at
Data Intelligence, Salesforce, 50 Fremont St. Suite 300, San Francisco, CA 94105, USA\\
\email{mohammad.ramezanali@gmail.com}
\and
P. P. Mitra  \at
Cold Spring Harbor Laboratory, 1 Bungtown Road, Cold Spring Harbor, NY 11734 USA\\
\email{mitra@cshl.edu}
\and
A. M. Sengupta \at
Department of Physics and Astronomy, Rutgers University, 136 Frelinghuysen Rd, Piscataway, NJ 08854 USA\\
%Center of Quantitative Biology, Rutgers University, 176 Frelinghuysen Rd, Piscataway, NJ 08854 USA\\
Center for Computational Biology, Flatiron Institute, 162 5\textsuperscript{th} Ave, New York, NY 10010, USA\\
\email{anirvans@physics.rutgers.edu}
}

\date{Received: date / Accepted: date}

\maketitle

\begin{abstract}
Recovery of an $N$-dimensional, $K$-sparse solution  $\mathbf{x}$ from an $M$-dimensional vector of measurements $\mathbf{y}$ for multivariate linear regression can be accomplished by minimizing a suitably penalized least-mean-square cost $||\mathbf{y}-\mathbf{H} \mathbf{x}||_2^2+\lambda V(\mathbf{x})$. Here $\mathbf{H}$ is a known matrix and  $V(\mathbf{x})$ is an algorithm-dependent sparsity-inducing penalty. For `random' $\mathbf{H}$, in the limit $\lambda \rightarrow 0$ and $M,N,K\rightarrow \infty$, keeping $\rho=K/N$ and $\alpha=M/N$ fixed, exact recovery is possible for $\alpha$ past a critical value $\alpha_c = \alpha(\rho)$. Assuming $\mathbf{x}$ has iid entries, the critical curve exhibits some universality, in that its shape does not depend on the distribution of $\mathbf{x}$. However, the algorithmic phase transition occurring at  $\alpha=\alpha_c$ and associated universality classes remain ill-understood from a statistical physics perspective, {\it i.e.} in terms of scaling exponents near the critical curve. In this article, we analyze the mean-field equations for two algorithms, Basis Pursuit ($V(\mathbf{x})=||\mathbf{x}||_{1} $) and Elastic Net ($V(\mathbf{x})= ||\mathbf{x}||_{1} +  \tfrac{g}{2} ||\mathbf{x}||_{2}^2$) and show that they belong to different universality classes in the sense of scaling exponents, with Mean Squared Error (MSE) of the recovered vector scaling as $\lambda^\frac{4}{3}$ and $\lambda$ respectively, for small $\lambda$ on the critical line. In the presence of additive noise, we find that, when $\alpha>\alpha_c$, MSE is minimized at a non-zero value for $\lambda$, whereas at $\alpha=\alpha_c$, MSE always increases with $\lambda$.


\end{abstract}

\section{Introduction}
\label{intro}
Variable or feature selection for multivariate linear regression is a classic problem in multivariate statistics. 
%Traditionally this has been typically accomplished by sequentially testing regression variables for goodness-of-fit with an assumed noise model, suitably penalized for the number of added regressors. Such procedures are "greedy" and do not necessarily yield global optima over the joint set of regressors. On the other hand, testing over all possible combinations of regressors is prohibitive as the number of such combinations grow exponentially with the number of regressors. This poses a particular challenge for modern machine learning applications where the number of putative independent variables can be large, often exceeding the number of measurements available. 
A popular approach to this problem is to perform the regression by including all possible predictors together with a sparsity-inducing penalty term in a joint optimization problem~\cite{tibshirani2015statistical}. In the "thermodynamic limit" of large data sets, where the number of measurements and number of predictors tend to infinity, these optimization problems show phase transition-like behavior. Namely, near a critical value of some important structural parameter, there is a qualitative change in the nature of solution of the penalized/regularized regression problem. Analysis of such algorithmic phase transitions forms the subject of this manuscript. 

An application of optimization employing sparsity-inducing penalties has been in compressed sensing where prior structure in the signals, in the form of sparsity in a suitable basis expansion, is exploited to reduce the number of measurements required to retrieve the signal. Early work in this area by Cand{\`e}s and Donoho~\cite{CandesCS,DonohoCS} exploited a combination of a convex relaxation of a non-convex penalty given by the number of non-zero coefficients, using an $\ell_1$-cost term, together with a random choice of measurement matrices to define the problem. A striking feature of this work was the computation of an algorithmic phase transition boundary separating a `good' regime in which perfect reconstruction is possible in suitable limits, from a `bad' regime where such reconstruction is impossible~\cite{Donoho_2009}.  

A typical statement of the sparse retrieval problem is an ill-posed linear equation, $\y=\H \x$ (noise free case), where $\y$ is an $M$ dimensional measurement vector, $\H$ is an $M \times N$ known measurement matrix, and $\x$ is an $N$ dimensional unknown parameter vector ($M < N$). Assume that $\y$ is generated by $\H\x_0$,  where $\x_0$ is the $N$ dimensional vector to be retrieved from the knowledge of $\y$ and $\H$. It is a priori known that  $\x_0$ has at most $K$ nonzero components. The task is to reconstruct this unknown vector. The ill-posedness of the underdetermined linear system is removed by imposing a sparsity constraint. For typical $\H$, as long as the number of unknowns, $K$, is less than the number of measurements $M$, these linear equalities have a unique sparse solution with high probability. 

A common formulation for the problem is to pose it as an optimization problem, defining $\hat \x(\lambda\sigma^2) ={\mathrm{arg\,min}}_\x\!\ \tfrac{1}{2\sigma^2}||\y-\H \x||_2^2 + \lambda\mathrm{V}$. In this sparse estimation framework, the purpose of the cost function $\mathrm{V}$ is to penalize the number of nonzero entries of $\x$ so that the sparsity property of the source is carried over to the solution $\hat \x$. The so-called $\ell_0$ norm, $||\x||_0=\lim_{p\rightarrow 0+}||\x||_p$,  where $||\x||_p=\sqrt[p]{\sum_a|x_a|^p}$ counts the number of nonzero elements of $\x$ (note that this is not a true norm as it does not satisfy the homogeneity condition).\\

Penalizing by the number of non-zero components leads to a non-convex optimization problem that is computationally hard because all possible combinations have to be tested. Chen et al~\cite{chen98} introduced the Basis Pursuit technique that uses the $\ell_1$- norm for enforcing sparsity as a computationally tractable convex relaxation of the original optimization problem and showed that the method provides correct results for sufficiently small $K$ (sufficient sparsity) under suitable assumptions. 
%The key idea is to replace the $\ell_0$ penalty by the $\ell_1$ norm penalty (i.e. $V= ||\x||_{1}$). 
%For the noisy case the resulting scheme is called LASSO~\cite{Tibshirani96} or Basis Pursuit denoising~\cite{chen98}. 

Another sparsity inducing cost combines the $\ell_1$ and $\ell_2$ norms, i.e. $\mathrm{V}(\x) = \lambda_1 ||\x||_{1} +  \tfrac{\lambda_2}{2} ||\x||_{2}^2$. The resulting optimization problem is known as the Elastic Net~\cite{Zou05}. $\ell_1$ and $\ell_2$ penalized regression methods both shrink the estimates of the regression coefficients towards zero to prevent overfitting due to (a) co-linearity of the covariates or (b) high-dimensionality. Although both penalties lead to shrinkage, namely, the regression coefficient estimates are ``shrunk" towards zero, the effects of $\ell_1$ and $\ell_2$ penalization are quite different. An $\ell_2$ penalty does not enforce strict sparsity and tends to result in all small but non-zero regression coefficients. In contrast, applying an $\ell_1$ penalty tends to result in some regression coefficients shrunk exactly to zero and other regression coefficients with comparatively little shrinkage. Combining $\ell_1$ and $\ell_2$ penalties tends to give a result in between, with fewer regression coefficients set to zero than in a pure $\ell_1$ setting, and more shrinkage of the other coefficients. The amount of shrinkage is determined by tuning parameters, $\lambda_1$ and $\lambda_2$. It was shown by Zou and Hasties~\cite{Zou05} that Elastic Net is effective at grouping highly correlated variables, i.e. they are either selected or removed from the model as a group.

In particular, for measurement matrices that have independent and identically distributed (iid) Gaussian entries, it has been shown that Basis Pursuit requires as low as $M>O(K\log(\N/K))$ measurements for perfect reconstruction~\cite{DonohoCS,CandesCS} with high probability. By now, there are a variety of approaches addressing this question. The original work of Donoho and Tanner~\cite{DonohoCS} addressed the problem where the coefficients are nonnegative using results on random projections of simplices in high dimensions~\cite{vershik1992asymptotic}. For the case of coefficients with unrestricted sign, analyses based on the message-passing method~\cite{DonohoAMP,bayati2011dynamics,bayati2012lasso}, and the replica formalism borrowed from statistical physics~\cite{Kabashima09,Ganguli10}, indicated that the $\ell_1$ norm minimization method and other analogous algorithms with polynomial time complexity exhibit a failure to retrieve the true solution at a sharp boundary as $N\rightarrow \infty$, with $M/N$ and $K/N$ being held fixed, analogous to a second-order (continuous) phase transition.%This is an algorithmic phase transition or zero-one law, where recovery fails or succeeds with high probability, the probability of correct recovery jumping from zero to one at the transition boundary.
Further work from multiple different angles provided rigorous ways of showing existence of a threshold. Gaussian process inequalities provided an alternative approach~\cite{stojnic2009various,stojnic2010ell_,stojnic2013rigorous}. Geometry of overlap between randomly rotated convex cones provided another~\cite{amelunxen2014living}. 

Several of the previous papers focus on the threshold and its relatively ``universal" nature (see, for example ~\cite{Donoho_2009}). For random Gaussian sensing matrices, introduction of correlation among entries can change such `universal' thresholds. However, in the literature of statistical physics, universality is usually associated with robustness of critical exponents (see below) to alterations in the original model setting, rather than with the threshold value for some parameter being universal~\cite{ma2018modern}.

In that spirit, in the current paper, we explore behavior near the performance transition boundary. Departures from the zero-noise or infinitesimal-regularization limit smooths out the sharp transition. In the statistical physics parlance, these are known as relevant perturbations. Critical exponents govern the power-law (occasionally logarithmic) dependence of performance metrics on the relevant perturbation strengths.    

Our approach of formulating an equivalent collection of single variable problems arises naturally from our earlier work on a two-step cavity method ~\cite{RMSCavity}. The cavity method results are essentially the consequences of self-consistency condition that are satisfied when belief propagation iterations converge~\cite{DonohoAMP,bayati2011dynamics,bayati2012lasso}. However, the final result of this approach is completely equivalent to replica computations~\cite{Kabashima09,Ganguli10} with the zero temperature limit taken carefully. We provide a sketch of the cavity method arguments in the appendix but relegate the details of the proof of the key propositions to a separate publication. In this paper, we focus on the applications of this method. Since second order transitions are classified by their critical exponents, we obtain these critical exponents associated with the effects of the strength of additive noise and of non-zero regularization parameter. Consequently we gain additional insights into the nature of a more general set of optimization problems.

\section{Notation and Paper Outline}
\label{sec:notation}

For matrices, we use boldface capital letters like $\H$, and we use $\H^{T}$, $\mathrm{tr(\H)}$,
to denote the transpose and trace, respectively. For vectors, we use boldface small letters like $\x$ with $x_a$ representing the $a^\mathrm{th}$ element of $\x$.
We use $\big[\dots\big]^{\mathrm{av}}_{\rm vars}$ to denote quenched averages, with the relevant quenched variables indicated in the subscript.  In particular, this average depends on two random variables $\x_{0}$ and $\H$ that are drawn from distribution $P_0(\x_{0})$ and $\mathcal{P}(\H)$. For a Gaussian random variable $\x$ with mean $\mu$ and variance $\upsilon$, we write the pdf as $\N(x; \mu, \upsilon)$ and, for the special case of $\N (x; 0, 1)$, we abbreviate the pdf as $\phi(x) = \frac{1}{\sqrt{2\pi}}\mathrm{e}^{-x^2/2}$ and write the cdf 
as $\Phi(x)= \int_x^\infty dz\,\phi(z)$. Dirac's delta function is written as $\delta(x)$ and $\delta_{mn}$ is the Kronecker delta symbol.

%where  $\Phi(\tau)= \int_\tau^\infty dz\, \phi(z)$, and $\phi(\tau) = \frac{1}{\sqrt{2\pi}}\mathrm{e}^{-\tau^2/2}$.

The rest of the paper is organized as follows. The problem formulation and Proposition~\ref{prop:minEeff}, following out of the cavity method~\cite{RMSCavity}, is given in Section~\ref{sec:formulation}. 
%This is also provided in more detail in Appendix~\ref{sec:recap}.  
We first treat the simple case of Ridge Regression~\cite{tikhonov1943stability}. Then, since the $\ell_{1}$-regularization is practically the most studied special case of the general setup, we apply the cavity approach and find a simple way to arrive at the two phases and the phase boundary in an insightful way and recover the known analytical formulation of results for Basis Pursuit~\cite{chen98}. Next, we study this phase transition boundary in various cases of additive-noise and non-zero trade-off parameter and find the behavior of the error as a function of these parameters and their scaling exponents at the phase boundary. Finally, we extend our analysis to the Elastic Net~\cite{Zou05} and obtain some new results. Conclusions and Summary are provided in Section~\ref{sec:summary}. The Appendix discusses origins of Proposition~\ref{prop:minEeff}  and also approaches Ridge Regression via Singular Value Decomposition.

\section{Problem Formulation and Methods}
\label{sec:formulation}

Consider the standard compressed sensing (CS) setup, $\y=\H\x_0+\bm\zeta$, where it is assumed that $\H \in {\mathfrak{R}}^{M \times N}$ represents the (known) measurement or design matrix $(M \leq N)$, the sparse vector $\x_0$ in ${\mathfrak{R}}^N$ is unknown , and the vector $\bm\zeta$ is  the measurement error with $E[\bm\zeta] = 0$, $E[\bm\zeta\bm\zeta^\prime]=\sigma_\zeta^2 \mathbf{I}_M$. Given the measurement matrix $\H$ and the measurements $\y$, the parameter vector $\x$ will be retrieved by minimizing the following penalized least mean squared cost. The estimated vector $\hat \x$ is defined by
\begin{equation}
\hat \x(\vartheta) = \underset{\x}{\argmin}\!\ \frac{1}{2\sigma^2}||\y-\H \x||_2^2 + \lambda \mathrm{V}(\x).
\label{eq:minV}
\end{equation}
$\vartheta=\lambda\sigma^2$ is a non-negative parameter giving relative weight between the first and second term in Eq.~\eqref{eq:minV} and  $V: \mathfrak{R}^{N} \rightarrow \mathfrak{R}$ a fixed non-negative regularization function or penalty term. Following literature practice we will focus on a $V(\x)=\sum_aU(x_a)$ that is convex and separable. In the sparse estimation framework, the best studied case is $U(x)=\lambda |x|$. It is known that for this penalty term, Eq.~\eqref{eq:minV} gives an exact reconstruction of $\x_0$ in a certain region of parameter space. 
%In this paper, our concern primarily is the noise-free case which is the choice $\vartheta \rightarrow 0^+$, i.e. a constrained optimization problem of minimizing $V(\x)$ subject to the constraint $\y=\H\x$. 
Using $\y=\H\x_0+\bm\zeta$, we write the minimization of Eq.~\eqref{eq:minV} as an optimization over the function $\E(\u)$ in terms of the error variable $\u=\x-\x_0$. 
\begin{equation}
\E(\u) = \frac{||\H\x_0 +\bm\zeta-\H\x||_2^2}{2 \sigma^2}+ V(\x) 
=\frac{1}{2\sigma^2} ||\H\u -\bm\zeta||_2^2  + V(\u+\x_0)
\label{eq:energy}
\end{equation}
Notice that the new cost function $\E(\u)$ itself is now a function of the input signal $\x_0$. In practice, $\x_0$ is not known and the original optimization problem has to be solved to obtain an estimate. However here we are interested in studying the statistical behavior of the minima of the optimzation problem over the distribution of instances of $\x_0$ and $\H$. For this purpose it is useful to reformulate the optimization problem as above, assume that $\x_0$ is fixed, and study the statistical behavior of the optimization problem for $\u$ in the presence of non-zero noise $\bm\zeta$ and different choices of $\H$. Eventually an average is taken over all distributions including that of $\x_0$ so the final results are not dependent on the choice of $\x_0$ or $\H$.  

The distribution of estimation error may be quantified by a suitable norm, for example $f(\x,\x_0)=\frac{1}{N}||\x-\x_0||_2^2=\frac{1}{N}||\u||_2^2$, as a measure of the inaccuracy of the reconstruction. The average of this quantity corresponds to the mean squared estimation error (MSE). 
\begin{equation}
{\rm MSE}\equiv \frac{1}{N}\big[||\x-\x_0||_2^2\big]^{\mathrm{av}}_{\x_0,\H,} = \frac{1}{N}\big[||\u||_2^2\big]^{\mathrm{av}}_{\x_0,\H}
\label{eq:mse}
\end{equation} 
Although in general $\H$ could be drawn from a non-Gaussian distribution, here we consider the special case in which $\H$ is Gaussian distributed. $\H^\mathrm{T}\H$ is nearly proportional to a unit matrix for a fixed choice of $\H$, and the first two moments of $\H$ satisfy
\begin{equation}
\big[H_{i a}\big]^{\mathrm{av}} = 0
\label{eq:gauss-av}
\end{equation} 
\begin{equation}
\big[ H_{i a} H_{j b}]^{\mathrm{av}} = \frac{1}{M}\delta_{i j}\delta_{a b}
\label{eq:gauss-covar}
\end{equation} 
The vector $\x_0$ is a random sample drawn from a factorized distribution $P_0(\x_0)=\prod_a p_0(x_{a0})$. Here we consider the sparsity inducing distribution $p_0(x_{a0})$ which has a continuous part and a delta function at origin: 
\begin{equation}
p_0(x_a)= \rho \pi(x_a)+ \left(1-\rho \right )\delta(x_a).
\label{eq:x0-dis}
\end{equation}
Now, we are ready to state the key proposition which allows us to do the further computations.
\begin{prop}[Equivalent Random Problem with Single Variable Optimizations]\label{prop:minEeff}
Let us consider the optimal solution $\tilde{\u}=\argmin_\u \E(\u)$ where $\E(\u)$ is defined as in \ref{eq:energy}, with $\x_0,\H,\bm\zeta$ chosen independently. The distribution of  $\x_0$ is given by \ref{eq:x0-dis}. $\H$ is a Gaussian random matrix satisfying \ref{eq:gauss-av} and \ref{eq:gauss-covar}. The components of $\bm\zeta$, $\zeta_a \overset{\text{iid}}{\sim}\mathcal N(0,\sigma_\zeta^2)$.  Then, the distribution of components of  $\tilde{\u}$, $\{\tilde{u}_a\}$ over the choice of $\H,\bm\zeta$, for fixed $\x_0$, would asymptotically be the same as the distribution of components $\{\hat u_a\}$, solving the following collection of single variable optimization problems with $\x_0$ held fixed but a random $\bm \xi$, and in the limit of $K,M,N\rightarrow\infty$ with $M/N=\alpha$ and $K/N=\rho$ held fixed.
\begin{align}
     & \hat u_a = \underset{u_a}{\argmin}\!\ \{ \frac{1}{2\seff} \left (u_a^2-2\xi_a u_a\right ) +  U(u_a+x_{0a}) \}\\
     & \xi_a \overset{\text{iid}}{\sim} \mathcal N( 0,\sigma_\xi^2) \,\,\,\mathrm{with} \,\,\, \sigma_\xi^2\equiv \sigma_\zeta^2 +\frac{q}{\alpha}  \label{eq:sigxi} \\
     & q \equiv \sum_a[\hat u_a^2]^{\mathrm{av}}_{x_0,\xi} \\
     &\sigma_{\mathrm{eff}}^2 \equiv \sigma^2+\frac{\overline{\chi}}{\alpha} \label{eq:sigeff-cavity} \quad    \\
   %  &\E_{\mathrm{eff}}(u_a;x_{0a},\xi_a)=\tfrac{1}{2\seff} \left (u _a^2-2\xi_a u_a\right ) +  U(u_a+x_{0a}).
    &\overline{\chi}\equiv\frac{1}{N}\sum_a\chi^{aa} \label{eq:chibar} 
 \end{align}
The quantity $q$ is the sum of the squared of error residuals, i.e. MSE. In addition, local susceptibility is obtained via $\hat u_a(f)-\hat u_a(0)=\chi^{aa}f_a $ with $f_a\rightarrow 0$ and $\hat u_a(f)$ is carried out by minimizing $\underset{u_a}{\mathrm{min}}\!\ \{ \frac{1}{2\seff} \left (u_a^2-2\xi_a u_a\right ) +  U(u_a+x_{0a})-f_a u_a \}$. In the end, summing over $\chi^{aa}$'s for all the instances of measurement matrix and then taking average over all nodes yields to the average local susceptibility, $\overline{\chi}$, and thus $\seff$. As it is presented in \cite{RMSCavity}, the asymptotic estimates of the local susceptibilities is given by
$$[\chi^{aa}(\x)]^{\mathrm{av}}=\bigg [U''(\hat{u}_{a}+x_{0a})+\frac{1}{\sigma_{\mathrm{eff}}^2}\bigg]^{-1}.$$
\end{prop}
Such self-consistent collections of random one-variable problems often arise in the physics of disordered systems~\cite{MPBook}. For an example in signal processing, where very similar conditions arise, see Xu and Kabashima's work on 1-bit compressed sensing~\cite{xu2013statistical}. In the next few sections, we solve the effective individual optimization in Proposition \ref{prop:minEeff} for the penalty function of the form $\lambda|x|^q$ with $q=1,2$. We point out the importance of $\overline{\chi}$ for distinguishing phases around the zero-temperature transition described by Donoho and Tanner~\cite{DonohoPT}. To facilitate further discussions, we summarize the symbols used in the next section in the Table \ref{tbl:symbols}.
\begin{table}[h]
\centering
\caption{Symbols used in this article}
\label{tbl:symbols}       % Give a unique label
\begin{tabular}{ll}
\hline\noalign{\smallskip}
Symbol      &Description \\
\noalign{\smallskip}\hline\noalign{\smallskip}
$u_a$ & Measure of residual error $x_a- x_{0a}$ \\
q & Mean squared error (MSE) \\
$\alpha$ & Measure for the number of constraints, $\frac{M}{N}$\\
$\rho$   &Measure for the sparsity, $\frac{K}{N}$ \\
$\lambda_1$ & $\ell_1$-norm regression coefficient   \\
$\lambda_2$ & $\ell_2$-norm regression coefficient   \\
$\sigma^2$ & Error variance on the constraint $\y=\H\x$ \\
$\vartheta $ &$\lambda\sigma^2 $\\
$\seff$ & Effective $\sigma^2$ given in the asymptotic limit of large $M, N$ \\
$\theta$ & $\lambda \seff$  \\
$\sxi$ & $\frac{q}{\alpha}$\\
$\sigma_\zeta^2$ & Variance of external noise\\
$\tau$ & $\frac{\theta}{\sigma_\xi}$ \\
\noalign{\smallskip}\hline
\end{tabular}
\end{table}

% \begin{table}
% \begin{center}
% \caption{Symbols}
% \begin{tabular}{ |p{2cm}|p{6.cm}|  }
% \hline
% \multicolumn{2}{|c|}{Symbols} \\
% \hline
% Symbol      &Description \\
% \hline
% $u_a$ & Measure of residual error $x_a- x_{0a}$ \\
% q & Mean squared error (MSE) \\
% $\alpha$ & Measure for the number of constraints, $\frac{M}{N}$\\
% $\rho$   &Measure for the sparsity, $\frac{K}{N}$ \\
% $\lambda_1$ & $\ell_1$-norm regression coefficient   \\
% $\lambda_2$ & $\ell_2$-norm regression coefficient   \\
% $\sigma^2$ & Error variance on the constraint $\y=\H\x$ \\
% $\vartheta $ &$\lambda\sigma^2 $\\
% $\seff$ & Effective $\sigma^2$ given in the asymptotic limit of large $M, N$ \\
% $\theta$ & $\lambda \seff$  \\
% $\sxi$ & $\frac{q}{\alpha}$\\
% $\sigma_\zeta^2$ & Variance of external noise\\
% $\tau$ & $\frac{\theta}{\sigma_\xi}$ \\

% \hline
% \end{tabular}
% \end{center}
% Table that presents symbols used in this article.
% \label{tbl:symbols}
% \end{table}

\section{Main Result}
\label{section main result}

\subsection{Ridge Regression}
\label{sec:ridge}
We warm up by using our method on the simplest from of regularization with $U(x)=\tfrac{\lambda}{2}x^2$, a penalty function that does not impose sparsity on the solutions. This is just a noise-free ridge regression with Tikhonov regularization~\cite{tikhonov1943stability}
\begin{equation}
\hat{\x}(\vartheta=\lambda\sigma^2)= \underset{\x}{\argmin} \{\frac{1}{2\sigma^2} || \H(\x-\x_0) ||_2^2 +\frac{\lambda}{2}||\x||_2^2\}.
\label{eq:ridge}
\end{equation}
The asymptotic limit of ridge regression has been previously studied by El Karoui~\cite{karoui2013asymptotic,el2018impact}. We could explicitly minimize $\x$ and proceed with our analysis using random matrix theory; however, we will apply first the self-consistency formalism we have developed (Proposition~\ref{prop:minEeff}).
\begin{equation}
\underset{u}{\mathrm{min}}\!\ \{ \frac{1}{2\sigma_\mathrm{ eff}^2}(u^2- 2\xi u) + \frac{\lambda}{2}(u+x_0)^2 \}
\label{eq:minEeff-ridge}
\end{equation}
Recalling that $u=x-x_0$ and identifying $\theta=\lambda \seff$, minimization of Eq.~\eqref{eq:minEeff-ridge} gives
\begin{equation}
\hat x=\frac{ x_0+\xi}{1+\theta}
\label{eq:min-ridge}
\end{equation}
This result can be used to determine $\sxi$ in Eq.~\eqref{eq:sigxi}
\begin{equation}
\sxi=\frac{q}{\alpha}=\frac{1}{\alpha}\big[u^2\big]^{\mathrm{av}}_{x_0,\xi}=\frac{\sxi+\theta^2\rho\big[x_0^2\big]^{\mathrm{av}}_{x_0}}{\alpha(1+\theta)^2}.
\label{eq:sxi-self-consistent-ridge}
\end{equation}
where $[\ldots]^{\mathrm{av}}_{x_0}$ means average over $\pi(x_0)$. 
One can see that with the ridge regression penalty function in \ref{prop:minEeff}, local susceptibility is the same everywhere: 
\begin{equation}
\overline{\chi}= \big[\lambda+\frac{1}{\sigma^2+\frac{\overline{\chi}}{\alpha}}\big]^{-1} \implies \theta = \left(\frac{1}{\lambda \overline{\chi} }-1\right)^{-1}.
\label{eq:chi-self-consistent-ridge}
\end{equation}
In particular in the $\vartheta \rightarrow 0$ limit, i.e. the minimal $\ell_2$ norm subject to linear constraints $\H\x=\H\x_0$, with $\lambda \overline{\chi}=1-\alpha$ gives $\theta=\alpha^{-1} -1 $. With the knowledge of $\theta$, the Eqs.~\eqref{eq:min-ridge}, and~\eqref{eq:sxi-self-consistent-ridge} lead us to
\begin{align}
\hat x=&\alpha( x_0+\xi)
\label{eq:min-ridge-extreme}\\ 
\sxi=&\frac{(1-\alpha)\rho}{\alpha}\big[x_0^2\big]^{\mathrm{av}}_{x_0}.
\label{eq:sxi-self-consistent-ridge-extreme}
\end{align}
which is the same conclusion from a formal singular value decomposition point of view (see Appendix~\ref{app:SVD}).
\begin{remark}
The estimated $x$ can be seen as a Gaussian variable, with $\alpha x_0$ as its mean and $ (1-\alpha)\alpha\rho[x_0^2]^{\mathrm{av}}_{x_0}$ as its variance. When the original variable $x_0=0$, we expect  the fluctuation  of the $\hat x$ around zero to be of the order $((1-\alpha)\alpha\rho[x_0^2]^{\mathrm{av}}_{x_0})^{1/2}$. We could set a threshold $\theta$ so that if $|\hat x|<\theta$ we truncate it to zero. We could then compute the false positive and false negative rates of such a procedure.  When $\rho<<\alpha/(1-\alpha)$, it is possible to choose a threshold $\theta$ such that $((1-\alpha)\alpha\rho[x_0^2]^{\mathrm{av}}_{x_0})^{1/2}<<\theta<<(\alpha[x_0^2]^{\mathrm{av}}_{x_0})^{1/2}$. With such a threshold, both error rates would be small. 
\end{remark}
%

%%%%%%%%%%%%%%%%%%%%%%%%%%%%%%%%%%%%%%%%%%%%%%%%%%%%%%%%%%%%%%%%%%%%%%%%%%%
%%%%%%%%%%%%%%%%%%%%%%%%%%%%%%%%%%%%%%%%%%%%%%%%%%%%%%%%%%%%%%%%%%%%%%%%%%%

\subsection{Basis Pursuit: $\ell_1$-norm Minimization}
\label{sec:l1}
In this section, we reconsider the much-analyzed case where the penalty function is the  $\ell_1$ norm of $\x$~\cite{DonohoPT,DonohoCS,CandesCS}. 
%Eq.~\eqref{eq:q},~\eqref{eq:dQ} and~\eqref{eq:Eeff-replica}
The reconstructed sparse solution is given by
\begin{equation}
\hat{\x}(\vartheta)= \underset{\x}{\mathrm{min}} \{\frac{1}{2\sigma^2} \left( \H(\x-\x_0)\right  )^2 + \lambda ||\x||_1\}.
\label{eq:basis-pursuit}
\end{equation}
Like in the case of ridge regression, we aim to solve the equations in proposition~\ref{prop:minEeff} for the potential $U(x)=\lambda |x|$ self-consistently. %equations \eqref{eq:sigxi} and \eqref{eq:sigeff} in the $\sigma^2  \rightarrow0$ limit. 
To determine $\theta$, once again we look at the local susceptibilities in \ref{prop:minEeff}. 
In this case $U''(x)$ is zero everywhere except at $x=0$, where it is formally infinite. Consequently, 
\begin{align}
\chi^{aa}&=0, \,\, \mathrm{ if }\,\,x_a=0\nonumber\\
\chi^{aa}&=\seff,\, \, \mathrm{ otherwise. } 
\end{align}
%The fact the $\chi^{aa}$ is the same for all non-zero values makes the analysis particularly simple. 
We define $\hat\rho$ to be the estimated sparsity, i.e. the fraction of $x_a$'s that are non-zero. Therefore $\overline{\chi}=\hat\rho\seff \,( \lambda \overline{\chi}= \hat \rho \theta)$ and 
\begin{equation}
\seff=\sigma^2+\frac{\overline{\chi}}{\alpha}=\sigma^2+\frac{\hat\rho\seff}{\alpha}
\end{equation}
implying
\begin{equation}
\theta(1-\frac{\hat\rho}{\alpha})=\vartheta
\label{eq:theta-rhohat}
\end{equation}
%Figure~\ref{fig:xvh} depicts the results of a computer experiment designed to estimate the susceptibility by linear programming as we vary the %external field $f$ applied on each individual node and compute the average of local displacement $\hat{x(f)}-\hat{x(0)}$ over all nodes. The slope %of this graph in the limit of $f\rightarrow0$ should give the value of susceptibility according to its definition in Eq.(30) in~\cite{RMSCavity}. This %experiment was performed using the CVXOPT package~\cite{Andersen10cvxopt}.
%Here we present four choices of $\alpha$ with one choice above $\alpha_c$ in the good regime, and three choices below $\alpha_c$ in the bad %regime. For the fixed dimension N = 200 and sparsity K=10, we reconstruct vector $x$ in the presence of external field $f$ according to the %following optimization 
%\begin{equation}
%\hat{\x}(\vartheta)= \underset{\x}{\mathrm{min}} \{\frac{1}{2\sigma^2} \left( \H(\x-\x_0)\right  )^2 + \lambda ||\x||_1
%-f u_{a}\}.
%\end{equation}
%while we subsample the standard normal measurement matrix $\H \in {\mathfrak{R}}^{50 \times 200}$ by decrement of 5. As it is shown, by i%ncreasing $f$, we find another near by solution $\hat{x(f)}$.
\begin{remark}\label{rmrk:PT}
The equation $\theta(1-\frac{\hat\rho}{\alpha})=\vartheta$ is central to understanding the $\vartheta  \rightarrow0$ limit and the associated phase transition. When $\vartheta$ goes to zero, we either have $\theta=0$ ($\hat\rho \neq \alpha$) or $\hat\rho=\alpha$ ($\theta \neq0$). These two conditions correspond to the two phases of the system, the first being the perfect reconstruction phase and the second, the non-zero error regime. In terms of average local susceptibility, the first phase has $\overline{\chi}=\hat\rho\theta=0$, while the second one has $\overline{\chi}\neq 0$.
\end{remark}

%Note the condition~\eqref{eq:theta-rhohat} involves $\sigma^2$ and $\seff$ explicitly, while the optimization result depends solely on $\vartheta=\lambda\sigma^2$ and not on $\sigma$ and $\lambda$ separately. However, just by multiplying the equation with $\lambda$, we could rewrite it as 
%\begin{equation}
%\theta(1-\frac{\hat\rho}{\alpha})=\vartheta
%\label{eq:theta-rhohat}
%\end{equation}
%with $\theta=\lambda\seff$. Further analysis of the self-consistency conditions depend crucially upon $\theta$. One could,  in principle, describe the system solely in terms of $\vartheta$ and $\theta$.

\begin{figure}[t]
\begin{center}
\includegraphics[width=.6\hsize]{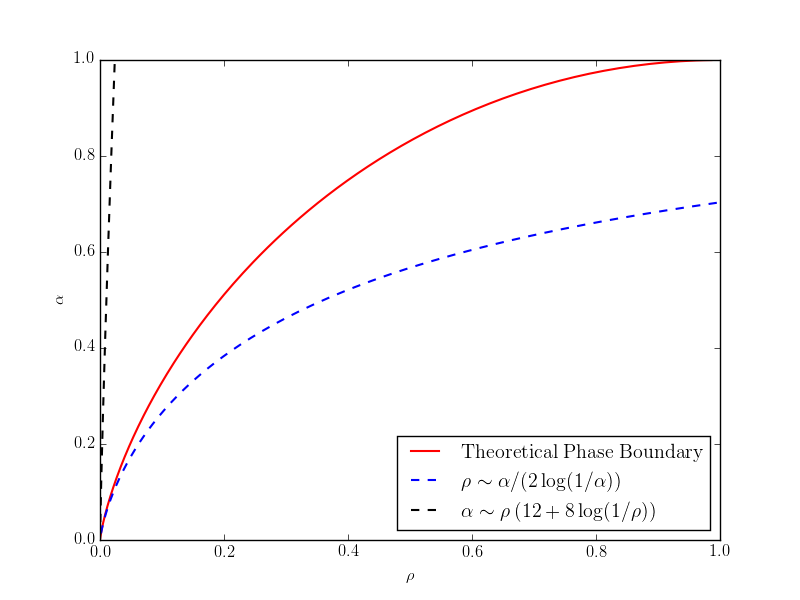}
\end{center}
\caption{The red curve is
the theoretical phase boundary obtained by solving Eq.~\eqref{eq:xierror} and~\eqref{eq:sefferror}. As $\rho\rightarrow 0$ this boundary is of the form $\rho=\alpha/(2\log(\frac{1}{\alpha}))$ as it is shown by dashed blue curve. The black dashed curve shows one of the restricted isometry property bounds \cite{rudelson2008sparse}. Perfect recovery occurs above the red curve. }
\label{fig:phase-boundary}
\end{figure}

Now we can set up the notation for the single variable optimization problem to find the value for $\sxi\,(\propto$ MSE) in these two regimes.
More precisely, by searching for the solutions to
\begin{equation}
\underset{u}{\mathrm{min}}\!\ \{ \frac{1}{2\sigma_\mathrm{ eff}^2}(u^2- 2\xi u) + \lambda |u+x_0| \}
\label{eq:minEeff-l1}
\end{equation}
we arrive at the following soft-thresholding function (also referred to as the proximal operator for the absolute value function) for the estimated value of $\hat x$ that we 
will denote by $\eta_{\mathrm{soft}}(t;\theta)$, with the variable $t = x_0+\xi$.
\begin{defn}[Soft Thresholding Function]
\begin{align}
\eta_{\mathrm{soft}}(t;\theta) = 
\begin{cases}
t-\theta& \mbox{ if $\theta\le t$,}\\
0& \mbox{ if $-\theta\le t\le \theta$,}\label{eq:soft-threshold}\\
t+\theta& \mbox{ if $ t<-\theta$.}
\end{cases}
\end{align}
\end{defn}
According to remark \ref{rmrk:PT}, the perfect reconstruction regime which ends to the phase boundary from above is the case where, as $\vartheta$ becomes small, $\theta$ becomes small as well.  From Eq.~\eqref{eq:soft-threshold}, there are three sources of error that can contribute to $\sxi$ in this regime (illustrated in Fig.~\ref{fig:soft-threshold}): 
\begin{enumerate}[a)] \item($x_0 \neq 0 \rightarrow \hat x = 0$) \\
Here $x_0$ was initially non-zero, but the estimated $\hat x$, due to the shift by $\xi$, has fallen into the $[-\theta, \theta]$ interval and then been truncated to zero. One can see that since $\theta$ is small, the probability of this event can be ignored for the time being%
\footnote{Under this circumstance, if $\xi$ remains of order one, then the error is dominated by $\xi$, i.e. $q (\mathrm{MSE})=\sxi$. However, this is not consistent with $\sxi=q/\alpha$, unless $\sxi=0$. Hence in this regime, we need to consider a $\sxi$ that is comparable to $\theta$. Therefore, as $\vartheta\rightarrow 0$,  we will have $\sxi\rightarrow 0$ and $q\rightarrow 0$, making the reconstruction perfect, i.e. the limit when $\vartheta,\theta,\sxi\rightarrow 0$ with $\tfrac{\theta}{\sigma_\xi}$ of order one.}.
%%%%%%%%%%%%%%%%%%%%%%%%%%%%%%%%%%%%%%%%%%%%%%%%%%%%%%%%%%%%%%%%%%%%%%%%%%%
\begin{figure}[t]
\begin{center}
\includegraphics[width=0.6\hsize]{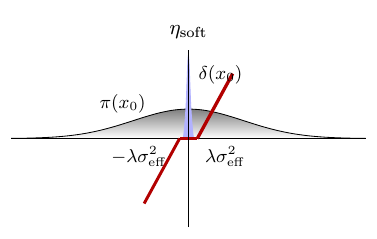}
\end{center}
\caption{The soft thresholding function ( in red) defined in~\eqref{eq:soft-threshold}. The non-zero entries of the sparse vector $\x_0$ drawn from random distribution is represented by $\pi$ ( in grey) and the zero components are represented by delta function (in blue)~\eqref{eq:x0-dis}}.
\label{fig:soft-threshold}
\end{figure}
%%%%%%%%%%%%%%%%%%%%%%%%%%%%%%%%%%%%%%%%%%%%%%%%%%%%%%%%%%%%%%%%%%%%%%%%%%%
\item ($x_0 \neq 0 \rightarrow \hat x \neq x_0$)\\
For non-zero $x_0$ that does not get set to zero, the contribution to MSE is
\begin{equation}
\rho[(\hat x-x_0)^2]^{\mathrm{av}}_{x_0,\xi} = \rho[\big(\xi-\theta\mathrm{sgn}(\hat x)\big)^2]^{\mathrm{av}}_{x_0,\xi} =\rho(\sxi+\theta^2)
\label{eq:xshift}
\end{equation}
\item ($x_0 = 0 \rightarrow \hat x \neq 0$) \\
Another source of error is the event when the $x_0$ is zero but $\hat x$ has fallen outside the interval  $[-\theta, \theta]$ and has been estimated to be non-zero. In this case, the contribution to MSE is 
\begin{align}
(1-\rho)[\hat x^2]^{\mathrm{av}}_{x_0,\xi} 
=&2(1-\rho)\int_{\theta}^{\infty}d\xi\,\dfrac{1}{\sqrt{2\pi\sxi}}\mathrm{e}^{-\frac{\xi^2}{2\sxi}}\big(\xi-\theta\big)^2\nonumber\\
=&2\sxi(1-\rho)\big\{(1+\tau^2)\Phi(\tau) - \tau \phi(\tau)\big\}
\label{eq:falsepositive}
\end{align}
 \end{enumerate} 
with $\tau=\tfrac{\theta}{\sigma_\xi}$. Adding up these contributions from Eq.~\eqref{eq:xshift} and~\eqref{eq:falsepositive}, we get the total MSE, $q$ (i.e. $\alpha\sxi$). Therefore using Eq.~\eqref{eq:sigxi}, $\sxi=q/\alpha$, and the knowledge of $\theta=0$ lead to the first parametric expression for the perfect reconstruction phase:
\begin{equation}
\alpha = 2(1-\rho)\big\{(1+\tau^2)\Phi(\tau) - \tau \phi(\tau)\big\}+ \rho(1+\tau^2).
\end{equation}
To determine $\hat \rho$, one can notice that if $x_0=0$, we have to have $|\xi|>\theta$ to lead to a non-zero $x$. On the other hand, since $\theta$ is small, a non-zero $x_0$ remains non-zero with probability approaching one. Counting all sources of the non-zero $\hat x$'s, then we have \footnote{Note that $\hat\rho>\rho$, even in the perfect reconstruction phase. That is because a fraction of $x_a$'s remain non-zero as long as $\vartheta>0$, and vanish only in the $\vartheta\rightarrow 0$ limit.} 
\begin{equation}
\hat\rho = 2(1-\rho)\Phi(\tau) + \rho.
\end{equation}
Recall that in the error-prone phase $\hat\rho=\alpha$ (Remark \ref{rmrk:PT}). This is due to the fact that $q,\sxi$ and therefore $\theta$ need to be non-zero in this regime. If the transition happens continuously, the condition for the phase boundary is $\alpha= \hat\rho=2(1-\rho)\Phi(\tau) + \rho$. Hence the relation between $\alpha$ and $\rho$ at the phase boundary is obtained by solving and eliminating $\tau$ from
     \begin{align}
     \alpha &= 2(1-\rho)\big\{(1+\tau^2)\Phi(\tau) - \tau \phi(\tau)\big\} + \rho(1+\tau^2)
     \label{eq:xierror}\\
     \alpha &= 2(1-\rho)\Phi(\tau) + \rho
     \label{eq:sefferror}
     \end{align}
Alternatively, Eq.~\eqref{eq:xierror} and~\eqref{eq:sefferror} can be solved for $\alpha$ and $\rho$ at the phase boundary and expressed parametrically as a function of $\tau$: %\begin{prop}[Phase Boundary Parametric Expression]
\begin{align}
&\alpha =\frac{2\phi(\tau)}{\tau + 2(\phi(\tau)- \tau\Phi(\tau))} \\
&\rho/\alpha  =1 - \frac{\tau \Phi(\tau)}{\phi(\tau)}
\end{align}
This leads to the phase diagram showing the transition from absolute success to absolute failure depicted in Fig.~\ref{fig:phase-boundary}. 
%\end{prop}
%%%%%%%%%%%%%%%%%%%%%%%%%%%%%%%%%%%%%%%%%%%%%%%%%%%%%%%%%%%%%%%%%%%%%%%%%%%
%%%%%%%%%%%%%%%%%%%%%%%%%%%%%%%%%%%%%%%%%%%%%%%%%%%%%%%%%%%%%%%%%%%%%%%%%%%
\begin{remark}
In the extremely sparse limit, $\rho<<1$, one can obtain a more explicit asymptotic relation between $\alpha$ and $\rho$. In this limit $\tau$ is large, and the dominant contributions are the second term, $\rho(1+\tau^2)$, from Eq.~\eqref{eq:xierror} and the first term, $2(1-\rho)\Phi(\tau)$, from Eq.~\eqref{eq:sefferror}. %\footnote{The important observation here is that this will still be valid even in the case of the correlated matrices.(\bf{add extra}}. 
Consequently,
\begin{align}
&\alpha  \approx \sqrt{\frac{2}{\pi}}\frac{e^{-\tfrac{\tau^2}{2}}}{\tau}\;  , \;\rho  \approx \sqrt{\frac{2}{\pi}}\frac{e^{-\tfrac{\tau^2}{2}}}{\tau^3} \\\nonumber
&\;\implies\;\rho \approx \frac{\alpha}{\tau^2}\; \approx \frac{\alpha}{2\log(1/\alpha)}
\end{align}
Therefore, in sparse limit, we have $\rho\sim \alpha/(2\log{\dfrac{1}{\alpha}})$ (see Fig. ~\ref{fig:phase-boundary}). Apart from a coefficient, this result has a similar form to the bounds from the restricted isometry property~\cite{Candes07}. From the figure it can be seen that this RIP bound is not very tight. 

\end{remark}
\section{Critical Exponents}
\label{sec:critical-exponent}

To get a better understanding of the nature of this phase transition and characterizing its behavior as one decreases $\alpha$ from above $\alpha_c(\rho)$ to below, we should search for solutions of the self-consistency equations in the error-prone regime where both $\theta$ and $\sxi$ remain O(1). In this case, we have to deal carefully with the possibility that $\hat x$ has been set to zero, because $x_0+\xi$ fell within $\pm \theta$. It is straightforward to show that the self-consistency equation for $\sxi$ becomes
\begin{align}
 \alpha &= \alpha \frac{\sigma_\zeta^2}{\sxi}+2(1-\rho)\big\{(1+\tau^2)\Phi(\tau) - \tau \phi(\tau)\big\}\nonumber\\
&+\rho\bigg[ \tau_0^2\big\{1 - \Phi(\tau+\tau_0)-\Phi(\tau-\tau_0))\big\}\nonumber \\
&+ (1+\tau^2)\big\{\Phi(\tau+\tau_0)+\Phi(\tau-\tau_0)\big\}\nonumber\\& - (\tau-\tau_0) \phi(\tau+\tau_0) - ( \tau+\tau_0) \phi(\tau-\tau_0)\bigg]^{\mathrm{av}}_{x_0}
\label{eq:sxi}
\end{align}
%\begin{align}
% \alpha &= 2(1-\rho)\big[(1+\tau^2)\Phi(\tau) - \tau \phi(\tau)\big] \nonumber\\
%&+\rho\bigg[ \frac{(\tau_+-\tau_-)^2}{4}\big\{1 - (\Phi(\tau_+)+\Phi(\tau_-))\big\}\nonumber \\
%&+ \big\{(1+\tau^2)(\Phi(\tau_+)+\Phi(\tau_-))\nonumber\\& - (\tau_- \phi(\tau_+) +  \tau_+ \phi(\tau_-))\big\}\bigg]^{\mathrm{av}}_{x_0}
%\label{eq:sxi}
%\end{align}
where $[\ldots]^{\mathrm{av}}_{x_0}$ means average over $\pi(x_0)$ and $\tau_0=\frac{x_0}{\sigma_\xi}$. The quantity $\tau$ and functions $\Phi(\tau)$ and $\phi(\tau)$ are defined as before. In addition, the parametric expression of Eq.~\eqref{eq:sefferror} becomes
\begin{equation}
\alpha = \frac{\vartheta}{\theta}+2(1-\rho)\Phi(\tau) + \rho\bigg[\Phi(\tau+\tau_0)+\Phi(\tau-\tau_0)\bigg]^{\mathrm{av}}_{x_0}
\label{eq:seff-lim}
\end{equation}
One should notice that in Eqs. \eqref{eq:sxi}, \eqref{eq:seff-lim} we included extra terms $\alpha \frac{\sigma_\zeta^2}{\sxi}$ coming from the additive noise and $\frac{\vartheta}{\theta}$ from not setting $\vartheta$ to zero, respectively. 

\begin{figure}[t]
\begin{center}
\includegraphics[width=.6\hsize]{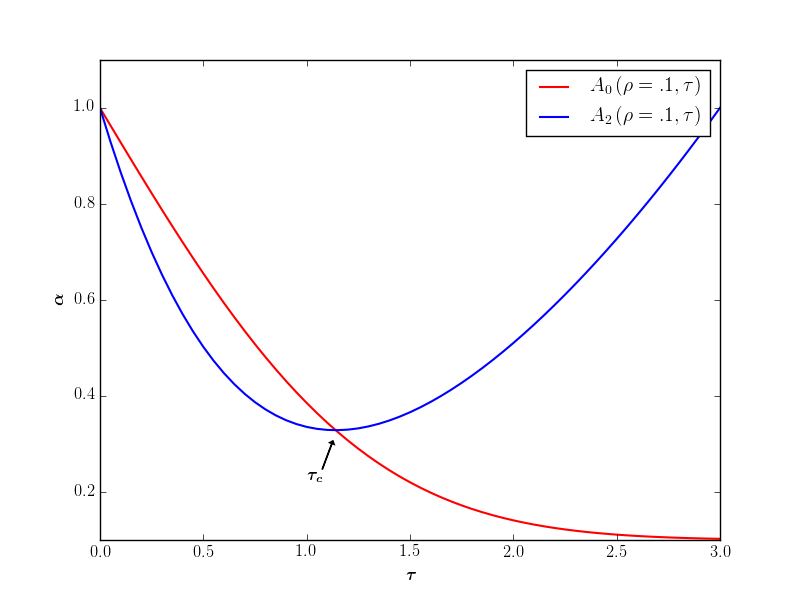}
\end{center}
\caption{The transition boundary is where the red and blue curves meet at the critical $\tau_c$.}
\label{fig:TauPlot}
\end{figure}

In order to better understand the behavior close to the transition where $\theta$ and $\sxi$ are small, we rewrite Eqs.~\eqref{eq:sxi} and~\eqref{eq:seff-lim} as \footnote{Note that, when $|\tau_0|=\tfrac{|x_0|}{\sigma_\xi}\rightarrow \infty$, $\Phi(\tau+\tau_0)+\Phi(\tau-\tau_0)\rightarrow 1$ and $(\tau-\tau_0) \phi(\tau+\tau_0),( \tau+\tau_0) \phi(\tau-\tau_0)\rightarrow 0$. The $\tau_0$ dependent expression inside $[\ldots]^{\mathrm{av}}_{x_0}$ in  Eq.~\eqref{eq:sxi} goes from $2\big\{(1+\tau^2)\Phi(\tau) - \tau \phi(\tau)\big\}$ to $1+\tau^2$ as $\tau_0$ goes from zero to infinity.  We wrote this expression as $1+\tau^2-\psi_\xi(\tau_0,\tau)$.}
\begin{align}
\alpha &=  \alpha \frac{\sigma_\zeta^2}{\sxi}+A_2(\rho,\tau) -\rho\big[\psi_\xi(\tau_0,\tau)\big]^{\mathrm{av}}_{x_0} 
\label{eq:sxi-rewrite}  \\
\alpha &= \alpha\frac{\vartheta}{\theta}+ A_0(\rho,\tau)  -\rho\big[\psi_{\theta}(\tau_0,\tau)\big]^{\mathrm{av}}_{x_0}
\label{eq:seff-rewrite}
\end{align}
where $\psi_\xi(\tau_0,\tau)$, $\psi_\theta(\tau_0,\tau)$ are even functions of $\tau_0$ that falls off quickly as $\tau_0$ becomes much larger than 1. For convenience we also defined
\begin{align}
A_2(\rho,\tau) &= 2(1-\rho)\big\{(1+\tau^2)\Phi(\tau) - \tau \phi(\tau)\big\} + \rho(1+\tau^2)\nonumber \\
A_0(\rho,\tau) &= 2(1-\rho)\Phi(\tau)+\rho
\label{eq:A20}
\end{align}
In Fig.~\ref{fig:TauPlot}, the behavior of these two functions are shown for a specific $\rho$. 
\begin{remark}\label{rmk:Tau-transition}
The transition boundary is where these two curves intersect at the point $\tau_c$. Note that $\frac{dA_2}{d\tau}=2\frac{A_2-A_0}{\tau}$. Thus, at the transition point $\tau_c$, $\frac{dA_2}{d\tau}=0$, i.e. $A_2$ behaves like $\sim\delta\tau^2$ ( $A_0$ goes as $\sim-\delta\tau$). As we will see in section \ref{sec:EN}, this relation will not be valid for Elastic Net. Therefore, we expect to have different critical behavior near the transition point for Elastic Net than Basis Pursuit.
\end{remark}
%%%%%%%%%%%%%%%%%%%%%%%%%%%%%%%%%%%%%%%%%%%%%%%%%%%%%%%%%%%%%%%%%%%%%%%%%%%
%%%%%%%%%%%%%%%%%%%%%%%%%%%%%%%%%%%%%%%%%%%%%%%%%%%%%%%%%%%%%%%%%%%%%%%%%%%
Moreover, we can write
\begin{align}
[\psi_\xi(\tau_0,\tau)]^{\mathrm{av}}_{x_0}=&\int dx_0\pi(x_0)\psi_\xi(\frac{x_0}{\sigma_\xi},\tau)\nonumber\\
=&\sigma_\xi\int d\tau_0\pi(\sigma_\xi\tau_0)\psi_\xi(\tau_0,\tau)
\end{align}
and get the same expression for $\psi_\theta(\tau_0,\tau)$. Thus, the small $\sxi$ behavior of these averages depends on how $\pi(x)$ behaves at small $x$. 
When $\pi(x)\sim F x^\gamma$ with $\gamma>-1$:
\begin{equation}
[\psi_\xi(\tau_0,\tau)]^{\mathrm{av}}_{x_0}\approx\sigma_\xi^{\gamma+1}\int d\tau_0\tau_0^{\gamma}\psi_\xi(\tau_0,\tau)\sim\sigma_\xi^{\gamma+1}
\end{equation}
Similarly 
$[\psi_{\theta}(\tau_0,\tau)]^{\mathrm{av}}_{x_0}\sim\sigma_\xi^{\gamma+1}$. Thus, the perturbations added to phase boundary Eqs.~\eqref{eq:xierror} and \eqref{eq:sefferror} are of the order of $\sigma_\xi^{\gamma+1}$.
Accordingly, in the case of a gapped distribution so that $\pi(x)=0$ when $|x|<\Delta$, we have:
\begin{equation}
[\psi_\xi(\tau_0,\tau)]^{\mathrm{av}}_{x_0}\approx\sigma_\xi \int_\Delta d\tau_0\psi_\xi(\tau_0,\tau)\sim e^{-\frac{\Delta^2}{\sigma_\xi^2}}\sigma_\xi
\end{equation}
And $[\psi_{\theta}(\tau_0,\tau)]^{\mathrm{av}}_{x_0}\sim e^{-\frac{\Delta^2}{\sigma_\xi^2}}\sigma_\xi$

\subsection{Into the Error-prone regime $ \big(\vartheta \rightarrow 0  \:\:\&\:\: \sigma_\zeta^2 = 0\big)$}

To find an estimate for the mean-squared error by entering into the error-prone regime, we express the phase boundary as $\alpha=\alpha_c(\rho), \tau=\tau_c(\rho)$ by solving Eqs.~\eqref{eq:xierror}, and \eqref{eq:sefferror}. To explore close to the phase boundary,  we can write $\alpha$ = $\alpha_c(\rho)-\delta\alpha$ and $\tau= \tau_c(\rho)-\delta\tau$. Since the perturbations to Eqs. \eqref{eq:xierror}, \eqref{eq:sefferror} for the case of $\pi(x)\sim F x^\gamma$ are of the the order $\sigma_\xi^{\gamma+1}$, from equation \eqref{eq:sxi-rewrite} we get
\begin{equation}
\delta\alpha \sim \sigma_\xi^{\gamma+1}= (\frac{q}{\alpha})^{\frac{\gamma+1}{2}}
\label{eq:critical-exponent}
\end{equation}
Therefore, for nonzero terms drawn from a distribution with nonzero density at the origin, Eq.~\eqref{eq:critical-exponent} tells us that 
the mean square error rises as
\begin{equation}
q(MSE)\sim (\alpha_c-\alpha)^{\frac{2}{\gamma+1}}
\end{equation}
%For Gaussian signals which $\gamma=0$, this is illustrated in Fig.~\ref{fig:MSE}.
Similarly, for $\pi(x)$ with a gap, we get a sharp rise for the error:
\begin{equation}
q\sim \frac{1}{\ln(1/(\alpha_c-\alpha))}
\end{equation}
\begin{remark}
The additional insight is that although the phase boundary $\alpha_c(\rho)$ does not depend on the distribution of non-zeros, the rise of the error does and becomes sharper when non-zero components are farther from zero. Moreover, the rise is continuous, i.e. it is a second-order phase transition and its critical exponent depends on the behavior of $\pi(x_0)$ near $x_0=0$. 
\end{remark}

\subsection{Role of an Additive Noise $ \big(\vartheta \rightarrow 0 \:\:\&\:\: \sigma_\zeta^2 \neq 0 \big)$} 
To examine the behavior of Eqs.~\eqref{eq:xierror}, and \eqref{eq:sefferror} close to the phase boundary within the presence of noise, once again, we Taylor expand them around the transition point where $\alpha=\alpha_c(\rho)$ and $\tau=\tau_c(\rho)$. Therefore, for the case of $\pi(x)\sim F x^{\gamma}$, Eqs. \eqref{eq:sxi-rewrite}, \eqref{eq:seff-rewrite} in terms of perturbing variables $\delta\alpha$ and $\delta\tau$ become:
\begin{align}
\delta\alpha = & \alpha_c \frac{\sigma_\zeta^2}{\sxi} + C \delta\tau^2 - D \sigma_\xi^{\gamma+1}+\cdots.
\label{eq:sxi-prtrb1} \\
\delta\alpha =& - C^\prime \delta\tau - D^\prime \sigma_\xi^{\gamma+1} +\cdots.
\label{eq:seff-prtrb1}
\end{align}
Where C, D, $C^\prime$ and $D^\prime$ are functions of $\rho$ and $\tau_c$, and `$\cdots$' contains higher order corrections. From Eq.~\eqref{eq:seff-prtrb1}, we have $\delta\tau = -(D^\prime/C^\prime) \sigma_\xi^{\gamma+1}$ which, by substitution to the first Eq.~\eqref{eq:sxi-prtrb1}, gives 
\begin{equation}
0 =  \alpha_c \frac{\sigma_\zeta^2}{\sxi} + \frac{C\,{D^\prime}^2}{{C^\prime}^2} \sigma_\xi^{2+2\gamma}  - D \sigma_\xi^{\gamma+1}   \nonumber \\
{\implies} \sxi \propto (\sigma_\zeta^2)^{2/(3+\gamma)}
\end{equation}
which we arrived at it by taking into account that $\sxi \rightarrow 0^+$.
With a similar calculation in the case with gapped distribution, we obtain
\begin{equation}
\sxi \propto \frac{1}{\ln(1/\sigma_\zeta^2)}
\end{equation}

\subsection{$\vartheta$ Trade-off in the Noisy system $ \big(\vartheta \neq 0 \:\:\&\:\: \sigma_\zeta^2 \neq 0 \big)$}
In the previous subsection, we considered the role of additive Gaussian noise in the behavior of the phase boundary near the transition from perfect reconstruction to the error regime. However, one should take into consideration that in most situations noise arises from several sources and there is no good estimation of either the level or distribution of the noise. Therefore, there is often a trade-off between the least squares of the residual and the $\ell_1$ norm of the solution. If the regularization is too much, the regularized solution does not fit the given signal properly as the residual error is too large. If the regularization is too small, the fit will be good but error will be more. One can control this trade-off and the sparsity of the solution by proper selection of the regularization parameter $\vartheta$. 
In the noise-free case, Taylor expansion of Eqs.~\eqref{eq:xierror}, and \eqref{eq:sefferror} close to the transition point leads to:
\begin{align}
\delta\alpha = &  C \delta\tau^2 - D \sigma_\xi^{\gamma+1}+\cdots
\label{eq:sxi-prtrb_theta} \\
\delta\alpha =&  \alpha_c\frac{\vartheta}{\theta} - C^\prime \delta\tau  - D^\prime \sigma_\xi^{\gamma+1} +\cdots
\label{eq:seff-prtrb_theta}
\end{align}
From Eq.~\eqref{eq:sxi-prtrb_theta}, we have $\delta\tau = (D/C)^{1/2} \sigma_\xi^{\frac{\gamma+1}{2}}$ which by substitution into Eq.~\eqref{eq:seff-prtrb_theta} and by taking into account that $\theta\sim\sigma_\xi$ gives 
\begin{equation}
0 =  \alpha_c\frac{\vartheta}{\theta}- \frac{C^\prime D^{1/2}}{C^{1/2}} \sigma_\xi^{\frac{\gamma+1}{2}}  - D \sigma_\xi^{\gamma+1}    \\
{\implies} \sxi \propto \vartheta^{\frac{4}{\gamma+3}}
\label{eq:xi-vtheta}
\end{equation}
 %This critical exponent is shown in Fig.~\ref{fig:tradeoff}. %Here we used the intersection points of the second derivates to determine the critical point since the finite size effects of the system were much reduced this way and we obtained a pretty good estimate of this critical exponent.
Similar calculation with the gapped distribution gives
\begin{equation}
\sxi \propto \frac{1}{\ln(1/\lambda)}
\end{equation}

As we mentioned earlier, a more interesting question would be that at what value of $\vartheta$, we will get the minimum error in the presence of noise. By adding noise to the system and expanding Eqs. \eqref{eq:seff-rewrite} and \eqref{eq:sxi-rewrite} in terms of perturbing variables $\delta\alpha$ and $\delta\tau$, we have 
\begin{align}
\delta\alpha = & \alpha_c \frac{\sigma_\zeta^2}{\sxi} + C \delta\tau^2 - D \sigma_\xi^{\gamma+1}+\cdots
\label{eq:sxi-prtrb_theta-noise} \\
\delta\alpha =&  \alpha_c\frac{\vartheta}{\theta} - C^\prime \delta\tau  - D^\prime \sigma_\xi ^{\gamma+1}+\cdots
\label{eq:seff-prtrb_theta-noise}
\end{align}
To have a solution, we get $\sigma_\xi^2\sim\vartheta^{\frac{2}{\gamma+2}}$ and $\sigma_\xi^2\sim(\sigma_\zeta^2)^{2/(\gamma+3)}$. Therefore, by tuning $\vartheta$ to $(\sigma_\zeta^2)^{\frac{\gamma+2}{\gamma+3}}$, the minimum error occurs. Similarly for the gapped non-zero distribution, $\sigma_\xi^2\sim \frac{1}{\ln(1/\vartheta)}$ and $\sigma_\xi^2\sim \frac{1}{\ln(1/\sigma_\zeta^2)}$. These scaling functions and critical exponents are summarized in the table \ref{tbl:critical-exponent-l1}.
\begin{table}
\centering
\caption{Critical exponents for $\ell_1$-norm minimization}
\label{tbl:critical-exponent-l1}       % Give a unique label
\begin{tabular}{ll}
\hline\noalign{\smallskip}
Input Variables    & Scaling Functions  \\
\noalign{\smallskip}\hline\noalign{\smallskip}
 \multicolumn{2}{c}{$\pi(x)\sim F x^\gamma$ with $\gamma>-1$} \\
 \hline
$\alpha\neq \alpha_c,\lambda \rightarrow 0, \sigma_\zeta^2=0$   & $MSE\sim (\alpha_c-\alpha)^{2/(1+\gamma)} $    \\ 
$\alpha=\alpha_c,\lambda \rightarrow 0, \sigma_\zeta^2\neq0$   &  $MSE\sim(\sigma_\zeta^2)^{2/(3+\gamma)} $   \\
  $\alpha=\alpha_c,\lambda \neq 0, \sigma_\zeta^2=0$    &  $MSE\sim\lambda^{4/(3+\gamma)}$      \\
  \hline
   \multicolumn{2}{c}{$\pi(x)=0$ for $|x|<\Delta$  } \\
 \hline
  $\alpha\neq\alpha_c,\lambda \rightarrow 0, \sigma_\zeta^2=0$   & $MSE\sim \frac{1}{\ln(1/(\alpha_c-\alpha))} $   \\ 
$\alpha=\alpha_c,\lambda \rightarrow 0, \sigma_\zeta^2\neq0$   &  $MSE\sim \frac{1}{\ln(1/\sigma_\zeta^2)} $    \\
  $\alpha=\alpha_c,\lambda \neq 0, \sigma_\zeta^2=0$    &  $MSE\sim\frac{1}{\ln(1/\lambda)}$        \\
\noalign{\smallskip}\hline
\end{tabular}
\end{table}
\subsection{Elastic Net}
\label{sec:EN}
As an application of our zero temperature cavity method, we consider how the phase transition is affected if we generalize the penalty function $V(\x)$ by adding a quadratic term $|\x|^2$ to the $\ell_1$ norm. This penalty function is used in the Elastic Net method of variable selection and regularization~\cite{Zou05}. The optimization problem becomes

\begin{equation}
\hat{\x}_\mathrm{EN} = \underset{\x}{\mathrm{min}}\{ \frac{1}{2\sigma^2} ||\y - \H\x||_2^2 + \lambda_1 ||\x||_1 + \frac{\lambda_2}{2} ||\x||_2^2\}
\end{equation}
In the noiseless reconstruction problem, $\y=\H\x_0$. We take the limit $\sigma^2\rightarrow 0$ and choose the distribution of $\H$ and $\x_0$ to be the same as in the previous sections.

Now $U''(x)=\lambda_2$ everywhere except at $x=0$, where it is formally infinite, leading to 
\begin{align}
\chi^{aa}&=0, \,\, \mathrm{ if }\,\,x_a=0\nonumber\\
\chi^{aa}&=\frac{\seff}{1+\lambda_2\seff},\, \, \mathrm{ otherwise. } 
\end{align}
Once more we define $\hat\rho$ to be fraction of $x_a$s that are non-zero. Then $\overline{\chi}=\frac{\hat\rho\seff}{1+\lambda_2\seff}$ and 
\begin{equation}
\seff=\sigma^2+\frac{\overline{\chi}}{\alpha}=\sigma^2+\frac{\hat\rho\seff}{\alpha(1+\lambda_2\seff)}
\end{equation}
implying
\begin{equation}
\seff\bigg\{1-\frac{\hat\rho}{\alpha(1+\lambda_2\seff)}\bigg\}=\sigma^2
\label{eq:sigeff-rhohat-elastic}
\end{equation}
In the $\sigma^2  \rightarrow0$ limit, the two phases are given by,  $\seff=0$  or $\hat\rho=\alpha(1+\lambda_2\seff)$. Again, the perfect reconstruction phase has $\overline{\chi}=\tfrac{\hat\rho\seff}{1+\lambda_2\seff}=0$ and  the  error-prone regime has $\overline{\chi}=\tfrac{\hat\rho\seff}{1+\lambda_2\seff}=\alpha\seff\neq 0$.

For the corresponding single variable optimization problem, we can still use the soft-thresholding function described in Eq.~\eqref{eq:soft-threshold}. The estimated value of $\hat x$ is once more given by $\eta_{\mathrm{soft}}(t;\theta)$, but with $t = \tfrac{x_0+\xi}{1+\lambda_2\seff}$ and $\theta =\tfrac{\lambda_1 \sigma_\mathrm{eff}^2}{1+\lambda_2\seff}$.
%%%%%%%%%%%%%%%%%%%%%%%%%%%%%%%%%%%%%%%%%%%%%%%%%%%%%%%%%%%%%%%%%%%%%%%%%%%
\begin{figure}
            \centering
                    \includegraphics[width=.6\hsize]{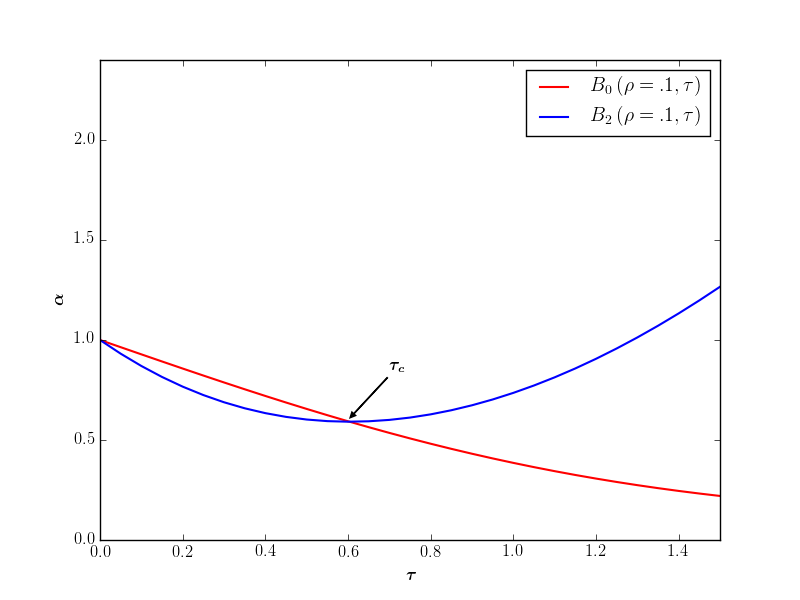}
                    \caption[The transition boundary is where the red and blue curves meet]{The transition boundary is where the red and blue curves meet at the critical $\tau_c$.  Unlike $\ell_1$-norm minimization, the slope at this point is not zero and there exists an additional linear term to the $B_2$ at the critical $\tau_c$ for Elastic Net. In the text we will see that this results in different critical behavior for Elastic Net.}

            \label{fig:TauPlotENet}
    \end{figure}

%%%%%%%%%%%%%%%%%%%%%%%%%%%%%%%%%%%%%%%%%%%%%%%%%%%%%%%%%%%%%%%%%%%%%%%%%%%
As before, we start in the perfect reconstruction phase, where 
 $\sigma^2,\seff,\sxi\rightarrow 0$ with $\tau=\tfrac{\lambda \seff}{\sigma_\xi}$ of order one. In this phase we ignore the case of non-zero $x_0$ leading to $\hat x=0$. The contribution to MSE for the non-zero $x_0$ is slightly different
\begin{align}
&\rho[(\hat x-x_0)^2]^{\mathrm{av}}_{x_0,\xi} = \rho\Bigg[\bigg(\frac{x_0+\xi-\lambda_1\seff\mathrm{sgn}(\hat x)}{1+\lambda_2\seff}-x_0\bigg)^2\Bigg]^{\mathrm{av}}_{x_0,\xi} \nonumber\\
\approx&\frac{\rho}{(1+\lambda_2\seff)^2}\bigg\{\sxi+(\lambda_1\seff)^2\big(1+\frac{\lambda_2^2}{\lambda_1^2}[x_0^2]^{\mathrm{av}}_{x_0}+\frac{\lambda_2}{\lambda_1}[|x_0|]^{\mathrm{av}}_{x_0}\big)\bigg\}
\label{eq:falsenegative-elastic}
\end{align}
The key approximation is that $[x_0\mathrm{sgn}(\hat x)]^{\mathrm{av}}_{x_0,\xi}\approx[|x_0|]^{\mathrm{av}}_{x_0}$, since in this limit typically $|\xi|<<|x_0|$ implying $\hat x$ and $x_0$ have the same sign.
The other source of error is the event when the $x_0$ is zero but $\hat x$ has fallen outside the interval  $[-\theta, \theta]$ and has been estimated to be non-zero. In this case, the contribution to MSE is 
\begin{align}
(1-\rho)[\hat x^2]^{\mathrm{av}}_{x_0,\xi} 
=&2(1-\rho)\int_{\lambda_1\seff}^{\infty}\dfrac{d\xi}{\sqrt{2\pi\sxi}}\mathrm{e}^{-\frac{\xi^2}{2\sxi}}\bigg(\frac{\xi-\lambda_1\seff}{1+\lambda_2\seff}\bigg)^2\nonumber\\
=&\frac{2\sxi(1-\rho)}{(1+\lambda_2\seff)^2}\big\{(1+\tau^2)\Phi(\tau) - \tau \phi(\tau)\big\}.
\label{eq:falsepositive-elastic}
\end{align}
Combining Eq.~\eqref{eq:falsenegative-elastic} and~\eqref{eq:falsepositive-elastic} in the self-consistency equation for $\sxi$ and remembering that $\sxi,\seff\rightarrow 0$ with $\tau=\tfrac{\lambda_1\seff}{\sigma_\xi}$ order one, we have 
\begin{align}
\alpha = &2(1-\rho)\big\{(1+\tau^2)\Phi(\tau) - \tau \phi(\tau)\big\}\nonumber\\
&+ \rho\bigg\{1+\tau^2\bigg(1+\frac{\lambda_2^2}{\lambda_1^2}[x_0^2]^{\mathrm{av}}_{x_0}+\frac{\lambda_2}{\lambda_1}[|x_0|]^{\mathrm{av}}_{x_0}\bigg)\bigg\}.
\end{align}

The equation for $\hat \rho$  remains the same in this limit.  The denominator $1+\lambda_2\seff$ does not matter for the thresholding condition. As a result once more
\begin{equation}
\hat\rho = 2(1-\rho)\Phi(\tau) + \rho.
\end{equation}
On the other hand, the condition for the phase boundary is
$\alpha= \hat\rho$. Thus, for the Elastic Net method, the phase boundary is obtained by solving and eliminating $\tau$ from
\begin{align}
\alpha &= 2(1-\rho)\big\{(1+\tau^2)\Phi(\tau) - \tau \phi(\tau)\big\}\nonumber\\
&+ \rho\bigg\{1+\tau^2\bigg(1+\frac{\lambda_2^2}{\lambda_1^2}[x_0^2]^{\mathrm{av}}_{x_0}+\frac{\lambda_2}{\lambda_1}[|x_0|]^{\mathrm{av}}_{x_0}\bigg)\bigg\}
\label{eq:xierror-elastic}\\
\alpha &= 2(1-\rho)\Phi(\tau) + \rho
\label{eq:sefferror-elastic}
\end{align}

In the case of Gaussian $\pi(x_0)$ with variance $\sigma^2_{x_0}$, the key dimensionless parameter is $\frac{\lambda_2\sigma_{x_0}}{\lambda_1}$, which determines the relative strength of the quadratic penalty term. It is important to note that unlike the $\ell_1$-norm minimization, the relation $\frac{dA_2}{d\tau}=2\frac{A_2-A_0}{\tau}$ in remark \ref{rmk:Tau-transition} does not hold for Elastic Net. Thus, Taylor expansion of Eq. \eqref{eq:xierror-elastic} (equivalent to the $A_2$ term in Eq. \eqref{eq:A20}) near the transition point has a linear contribution with positive slope as well as quadratic one (See Fig. \ref{fig:TauPlotENet}). The theoretical critical exponents can be derived in the same way as described in section \ref{sec:critical-exponent}. We only mention the results in the table \ref{tbl:critical-exponent-enet}.

\begin{table}
\centering
\caption{Critical exponents for the Elastic Net}
\label{tbl:critical-exponent-enet}       % Give a unique label
\begin{tabular}{ll}
\hline\noalign{\smallskip}
Input Variables    & Scaling Functions  \\
\noalign{\smallskip}\hline\noalign{\smallskip}
 \multicolumn{2}{c}{$\pi(x)\sim F x^\gamma$ with $\gamma>-1$} \\
 \hline
 $\alpha\neq\alpha_c,\lambda \rightarrow 0, \sigma_\zeta^2=0$   & $MSE\sim (\alpha_c-\alpha)^{2/(1+\gamma)} $    \\ 
$\alpha=\alpha_c,\lambda \rightarrow 0, \sigma_\zeta^2\neq0$   &  $MSE\sim(\sigma_\zeta^2)^{2/(3+\gamma)} $   \\
  $\alpha=\alpha_c,\lambda \neq 0, \sigma_\zeta^2=0$    &  $MSE\sim\lambda^{2/(2+\gamma)}$      \\
  \hline
   \multicolumn{2}{c}{$\pi(x)=0$ for $|x|<\Delta$  } \\
 \hline
  $\alpha\neq\alpha_c,\lambda \rightarrow 0, \sigma_\zeta^2=0$   & $MSE\sim \frac{1}{\ln(1/(\alpha_c-\alpha))} $   \\ 
$\alpha=\alpha_c,\lambda \rightarrow 0, \sigma_\zeta^2\neq0$   &  $MSE\sim \frac{1}{\ln(1/\sigma_\zeta^2)} $    \\
  $\alpha=\alpha_c,\lambda \neq 0, \sigma_\zeta^2=0$    &  $MSE\sim\frac{1}{\ln(1/\lambda)}$        \\
\noalign{\smallskip}\hline
\end{tabular}
\end{table}
\section{Numerical Experiments}
\label{sec:numeric}

\begin{figure}[t]
\begin{center}
\includegraphics[width=.6\hsize]{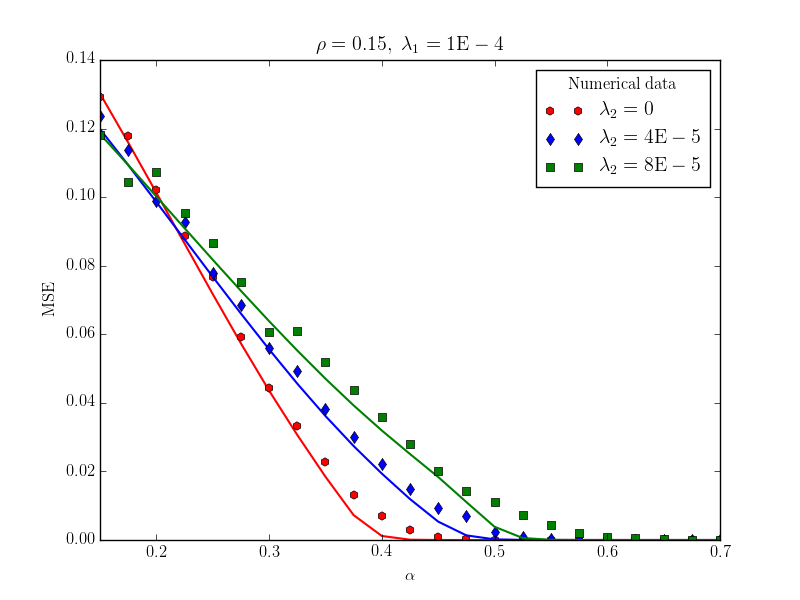}
\end{center}
\caption{Comparison of MSE for different $\lambda_{2}$. Each solid curve represents the theoretical estimate for MSE as described in Sec.~\ref{sec:l1} and Sec.~\ref{sec:EN}.  Numerical data for different $\lambda_{2}$ is shown with the markers. We use CVXOPT quadratic programming to find MSE for 3 values of $\lambda_{2}/\lambda_1$: 0, 0.4, 0.8. Notice that, for the Elastic Net ($\lambda_2\neq0$, the transition happens at higher $\alpha$ compared to $\ell_1$-norm minimization.  }
\label{fig:MSE-ENet}
\end{figure}

This section describes the numerical implementation for examining
critical exponents that we obtained in section \ref{sec:critical-exponent} and comparison with some numerical experiments. 
First, we compute MSE for $\ell_1$-norm minimization and Elastic Net (see Fig. \ref{fig:MSE-ENet}).
The matrix $\H$ is obtained by first filling it with independent
samples of a Gaussian distribution with variance $1/M$.

%%%%%%%%%%%%%%%%%%%%%%%%%%%%%%%%%%%%%%%%%%%%%%%%%%%%%%%%%%%%%%%%%%%%%%%%%%%
%%%%%%%%%%%%%%%%%%%%%%%%%%%%%%%%%%%%%%%%%%%%%%%%%%%%%%%%%%%%%%%%%%%%%%%%%%%

In this example,
$N = 200$, $K = 30$, the original signal $\x$ contains 30 randomly placed elements driven from a standard Gaussian distribution, i.e. $\gamma=0$. The numerical experiment is carried out using CVXOPT quadratic programming \cite{Andersen10cvxopt} and for $\lambda_1=\mathrm{1E-8}$ and $\lambda_2=0,  .4, .8$ of $\lambda_1$ (to relate with compressed sensing literature, we have set $\sigma^2=1$, i.e. $\vartheta=\lambda$). As it can be seen, the reconstruction error exhibits a slightly higher mean squared error (MSE) with respect to the theoretical result. We believe that effect is due to the finiteness of $M,N$ and $K$.
%%%%%%%%%%%%%%%%%%%%%%%%%%%%%%%%%%%%%%%%%%%%%%%%%%%%%%%%%%%%%%%%%%%%%%%%%%%
%\begin{figure}[t]
%\begin{center}
%\includegraphics[width=1.0\hsize]{MSE}
%\end{center}
%\caption{ Comparison of numerical MSE with  the theoretical result. The solid curve represents the theoretical estimate for MSE (see section~\ref{sec:l1} ). Numerical data is shown with the markers. The blue markers represents the contribution to MSE from the zero components of the signal. The green markers represents the contribution to MSE from the non-zero sparse components of the signal. }
%\label{fig:MSE}
%\end{figure}
%%%%%%%%%%%%%%%%%%%%%%%%%%%%%%%%%%%%%%%%%%%%%%%%%%%%%%%%%%%%%%%%%%%%%%%%%%%

\par
\begin{figure}[t]
\begin{center}
\includegraphics[width=.6\hsize]{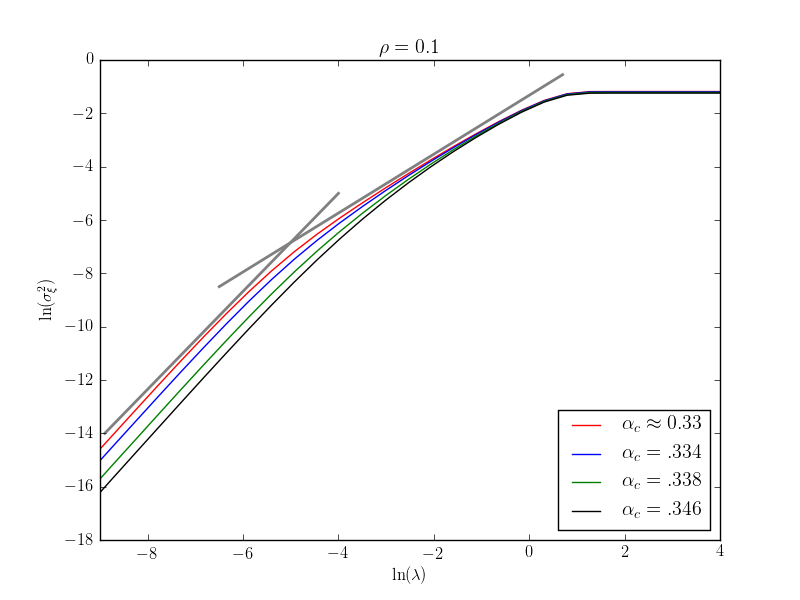}
\end{center}
\caption{Following the trends where the curves merge, we can find the critical exponent near phase transition. The gray lines show two different exponents near the transition going from the slope 1.33 to 2.}% at the intersection points of the second derivates shown in the subplot
\label{fig:tradeoff}
\end{figure}

 Next, we confirm the exponent in Eq. \eqref{eq:xi-vtheta} by plotting the theoretical expression in Eq. \eqref{eq:sxi-rewrite}. This is shown in Fig. \ref{fig:tradeoff}.
%%%%%%%%%%%%%%%%%%%%%%%%%%%%%%%%%%%%%%%%%%%%%%%%%%%%%%%%%%%%%%%%%%%%%%%%%%%
%%%%%%%%%%%%%%%%%%%%%%%%%%%%%%%%%%%%%%%%%%%%%%%%%%%%%%%%%%%%%%%%%%%%%%%%%%%

%%%%%%%%%%%%%%%%%%%%%%%%%%%%%%%%%%%%%%%%%%%%%%%%%%%%%%%%%%%%%%%%%%%%%%%%%%%
\begin{figure}[t]
\begin{center}
\includegraphics[width=.6\hsize]{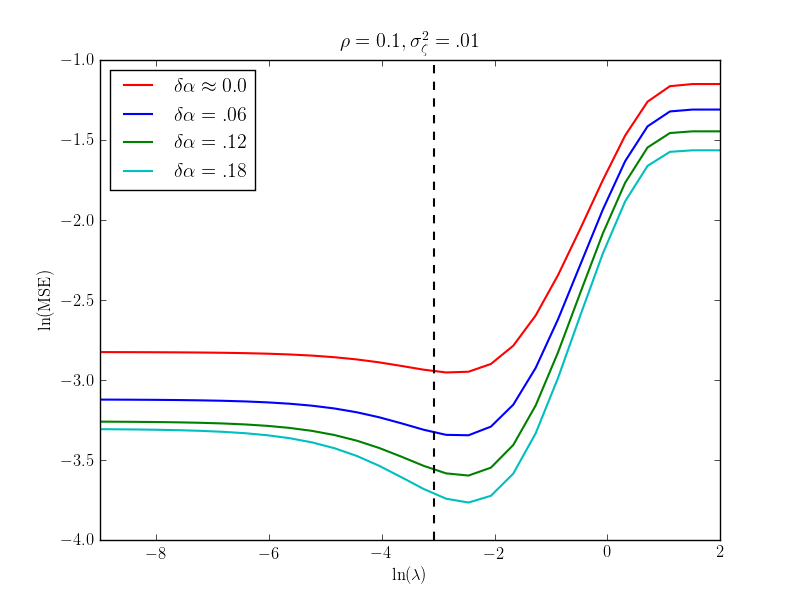}
\end{center}
\caption{Varying $\lambda$ sweeps out entire optimal tradeoff curves. The vertical black dashed line is located at $\ln(\lambda)=\frac{2}{3}\ln(\sigma_\zeta^2)$ in which the theoretical minimum error near phase transition occurs.}
\label{fig:tradeoff1}
\end{figure}
%%%%%%%%%%%%%%%%%%%%%%%%%%%%%%%%%%%%%%%%%%%%%%%%%%%%%%%%%%%%%%%%%%%%%%%%%%%

In the end, we consider the important case where the external noise is non-zero and we are looking for a trade-off for $\lambda$ where the reconstruction error is minimized. Once more, using Eq. \eqref{eq:sxi-rewrite}, this is shown in Fig. \ref{fig:tradeoff1}. One can see that having a non-zero $\lambda$ does not significantly help with the error in recovery of the signal at the transition point for $\alpha$ (the red curve). However, further from the transition line, the non-zero value of $\lambda$ can result in lower MSE. It should be noted that the dependence of the MSE on $\lambda$ is fairly shallow, so that letting $\lambda$ go to zero does not significantly increase the MSE. This is of interest since such a choice obviates the need to fix the regularization parameter.

\section{Conclusion and Summary of Results}
\label{sec:summary}

We have presented a different approach to the study of the statistical properties of sparsity-penalized multivariate linear regression, and compressed sensing problems, compared to the standard replica formalism and message-passing algorithms. We exploited a local susceptibility to understand the simple case of Ridge Regression and then to find a simple method for deriving the phase boundary known for the Basis Pursuit. We showed that this transition is continuous (second order) and analyzed the critical behaviors, including scaling functions and critical exponents that are uniquely determined by the universality class of the phase transition. Our considerations demonstrate that the Basis Pursuit and Elastic Net algorithms belong to different universality classes in the   usual statistical physics sense. It would be interesting to extend these considerations to other algorithms that may also be studied using the cavity mean field approach. 

We have stressed the important role of a local error susceptibility introduced by the zero-temperature cavity method as a powerful tool in sparse recovery problems. It turns out that the perfect reconstruction phase corresponds to vanishing average local susceptibility, indicating that the solution of the optimization problem has an underlying robustness to perturbations in this phase. We expect that the structure of the error susceptibility enjoys unique properties and applicability beyond the standard sparse setting traditionally considered in compressed sensing.

%We postpone the derivation of the modified self-consistency equations in the presence of noise and the correlated measurement matrices to our future work.
%

\begin{acknowledgements}
This work was supported by the National Science Foundation INSPIRE (track 1) award 1344069.
Part of this paper was written while two of the authors (AMS and MR) were visiting Center for Computational Biology at Flatiron Institute. We are grateful for their hospitality. 
\end{acknowledgements}

% \section*{Acknowledgment}
% %\addcontentsline{toc}{section}{acknowledgments}
% This work was supported by the National Science Foundation INSPIRE (track 1) award 1344069.
% The final version of this paper was written while two of the authors (AMS and MR) were visiting Simons Center for Data Analysis at Simons Foundation. We are grateful for their hospitality. 
% %\end{acknowledgments}

%\appendix
%\section*{Appendix}
\begin{appendices}
\section{Equivalent Single Variable Optimization Problem}
Here, we provide a sketch of our cavity argument. A detailed derivation will be left for the published version of our preprint~\cite{RMSCavity}. From an algorithmic point of view, the cavity method is related to message passing algorithms, but we will assume that the algorithm convergences on a state. We wish to find an approximate statistical description of that state. Once we are done discussing cavity method, we also briefly mention how to connect these results to the replica calculations found in~\cite{Kabashima09,Ganguli10}. 

We will consider the case where the function $V$ is twice differentiable. To construct potentials like the $\ell_1$ norm, we use second differentiable functions like $r\ln(2\cosh(x/r))$, which tends to $|x|$ when $r$ goes to zero. We can study the solution for $r>0$ and then take the appropriate limit.

Minimization of the original penalized regression objective function is mathematically equivalent to the minimization of $\E(\u)$ over $\u$ (even if, in practice, we do not know the explicit form of $\E(\u)$). 
\begin{align}
&\min_\u\E(\u) \nonumber\\
=&\min_\u\frac{1}{2\sigma^2} ||\H\u -\bm\zeta||_2^2  + V(\u+\x_0)\nonumber\\
=&\min_\u\max_\z -\frac{\sigma^2}{2} ||\z||_2^2+\z^T(\H\u -\bm\zeta) + V(\u+\x_0)\nonumber\\
=&\min_\u\max_\z - \sum_{i=1}^M\big(\frac{\sigma^2}{2}z_i^2-\zeta_iz_i\big)-\sum_{i=1}^M\sum_{a=1}^Nz_iH_{ia}u_a + \sum_{a=1}^NU(u_a+x_{0a})
\label{eq:dual_energy}
\end{align}

Note that the $z_i$ variables and the $u_a$ variables only interact via the random measurement matrix $\H$. From this point on, our arguments are similar to that of Xu and Kabashima~\cite{xu2013statistical}. We try to find single variable functions $\E_a(u_a)$ and $\E_i(z_i)$ whose optimization mimics the full optimization problem.

We consider the problem with an $a$-cavity, meaning, a problem where the variable $x_a=u_a+x_{0a}$ has been set to zero. We also consider a problem with an $i$-cavity, namely a problem, where $z_i$ has been set to zero. The single variable functions are constructed by introducing the inactive/missing variable into the corresponding cavity. The discussion becomes simpler if we assume the optimization in the systems with cavities are already well-approximated by optimizing over sum of single variable functions of the following forms (as is done in Xu and Kabashima~\cite{xu2013statistical}).
\begin{equation}
    \E_a(u_a)=\frac{A_a}{2}u_a^2-F_au_a+U(x_{0a}+u_a)
\end{equation}
and
\begin{equation}
\E_i(z_i)=-\frac{B_i}{2}z_i^2+K_iz_i.
\end{equation}
For simplicity, we will call these single variable functions potentials.

We will set up slightly more involved notation for these parameters for a cavity system. With $a$ missing, the potentials for $z_i$ are represented by
\begin{equation}
\E_{i\rightarrow a}(z_i)
=-\frac{B_{i\rightarrow a}}{2}z_i^2+K_{i\rightarrow a}z_i.
\label{eq:param-eia}
\end{equation}
Similarly, with $i$ missing, we have
\begin{equation}
    \E_{a\rightarrow i}(u_a)=\frac{A_{a\rightarrow i}}{2}u_a^2-F_{a\rightarrow i}u_a+U(x_{0a}+u_a).
    \label{eq:param-eai}
\end{equation}
We argue that the corresponding parameters with or without cavity are nearly the same.

\subsection*{Step 1: Introducing $x_a=$ to the $a$-cavity}
\begin{equation}
\E_a(u_a)=V(x_{0a}+u_a)
+\sum_{i=1}^M\big\{\max_{z_i} (-z_iH_{ia}u_a+\E_{i\rightarrow a}(z_i))\big\} 
\end{equation}

\begin{equation}
\E_{a\rightarrow i}(u_a)=V(x_{0a}+u_a)
+\sum_{j\neq i}\big\{\max_{z_j} (-z_jH_{ja}u_a+\E_{i\rightarrow a}(z_j))\big\} 
\end{equation}

From here, using the parametrization of $\E_{i\rightarrow a}(z_i)$ according to Eq.~\ref{eq:param-eia}, we know that optimal $z_j=\tfrac{H_{ia}u_a-K_{j\rightarrow a}}{K_{j\rightarrow a}}$, leading to
\begin{align}
A_a&=\sum_{i=1 }^M\frac{H_{ia}^2}{B_{i\rightarrow a}}
\label{eq:Aa}
\\
F_a&=-\sum_{i=1 }^M\frac{H_{ia}K_{i\rightarrow a}}{B_{i\rightarrow a}}
\label{eq:Fa}
\end{align}
and
\begin{align}
A_{a\rightarrow i}&=\sum_{j\neq i}\frac{H_{ja}^2}{B_{j\rightarrow a}}\\
F_{a\rightarrow i}&=-\sum_{j\neq i}\frac{H_{ja}K_{j\rightarrow a}}{B_{j\rightarrow a}}.
\end{align}

Since $H_{ia}\sim \tfrac{1}{\sqrt{N}}$, $A_a\approx A_{a\rightarrow i}$ and $F_a\approx F_{a\rightarrow i}$.

\subsection*{Step 2: Introducing $z_i$ to the $i$-cavity}
\begin{equation}
\E_i(z_i)=-\frac{\sigma^2}{2}z_i^2
+\zeta_iz_i
+\sum_{a=1}^N\big\{\min_{u_a} (-z_iH_{ia}u_a+\E_{a\rightarrow i}(u_a))\big\} 
\end{equation}

\begin{equation}
\E_{i\rightarrow a}(z_i)=-\frac{\sigma^2}{2}z_i^2
+\zeta_iz_i
+\sum_{b\neq a}\big\{\min_{u_b} (-z_iH_{ib}u_b+\E_{i\rightarrow a}(u_b))\big\} 
\end{equation}

Now, we use the parametrization of $\E_{a\rightarrow i}(u_a)$ according to Eq.~\ref{eq:param-eai}, and get the optimal $u_b$ satisfies $$-z_iH_{ib}+A_{b\rightarrow i}u_b-F_{b\rightarrow i}+U'(x_{0a}+u_a)=0.$$
Since $z_iH_{ib}\sim \tfrac{1}{\sqrt{N}}$, we can expand this equation around $z_i=0,\bar u_b$ and find that 

\begin{align}
B_i&=\sigma^2+\sum_{a=1 }^N\frac{H_{ia}^2}{A_{i\rightarrow a}+U''(x_{0a}+u_a)}
\label{eq:Bi}
\\
K_i&=\zeta_i-\sum_{a=1}^NH_{ia}\bar u_a
\label{eq:Ki}
\end{align}

We could derive equations for $B_{i\rightarrow a}$ and $K_{i\rightarrow a}$, like before, but we know that we can ignore the difference between these parameters and $B_i$ and $K_i$ respectively. Also, we have optimal $u_a\approx\bar u_a$.

\subsection{Step 3: Putting it all together}

We now only deal with $A_a,B_i$ etc. From Eq.~\ref{eq:Aa} and Eq.~\ref{eq:Bi}
\begin{align}
A_a&=\sum_{i=1 }^M\frac{H_{ia}^2}{B_i}\\
B_i&=\sigma^2+\sum_{a=1 }^N\frac{H_{ia}^2}{A_i+U''(x_{0a}+u_a)}
\end{align}
For large $M,N$ the expressions are self-averaging. One can essentially replace $H_{ia}^2$ by $\tfrac{1}{M}=\tfrac{1}{\alpha N}$ and see that $A_a=A,B_i=B$. In other words, these quantities are essentially index independent.

\begin{align}
A&=\frac{1}{B}\\
B&=\sigma^2+\frac{1}{\alpha N}\sum_{a=1 }^N\frac{1}{A+U''(x_{0a}+u_a)}
\end{align}

If we identify $B=\seff=\tfrac{1}{A}$, we see that we got 
\begin{equation}
    \seff=\sigma^2+\frac{\bar \chi}{\alpha}
\end{equation}
according to the definition in Proposition~\ref{prop:minEeff}.

To get our final result, we need $F_a$. Using Eq.~\ref{eq:Fa} and Eq.~\ref{eq:Ki} and the various approximations
\begin{equation}
F_a=-\sum_{i=1 }^M\frac{H_{ia}K_{i\rightarrow a}}{B_{i\rightarrow a}}\approx-\frac{1}{B}\big[\sum_{i=1}^M(\zeta_i-\sum_{b\neq a}H_{ib}u_b)H_{ia}\big]\equiv-\frac{\xi_a}{\seff}.
\end{equation}
We, thus, identify the term inside the square bracket as $\xi_a$. Its distribution over different choices of $\H$ and $\bm\zeta$,is approximately normal with mean zero and variance $=\sigma_\zeta^2+\tfrac{\alpha}{N}\sum_{b\neq a}u_b^2\approx\sigma_\zeta^2+\alpha q.$ Also, $\xi_a$ and $\xi_b$ are nearly uncorrelated for $a\neq b$.

At the end we get
\begin{equation}
 \E_a(u_a)=\frac{A_a}{2}u_a^2-F_au_a+U(x_{0a}+u_a)=\frac{1}{2\seff}u_a^2-\frac{\xi_au_a}{\seff}+U(x_{0a}+u_a)
 \end{equation}
 which is the same expression as in Proposition~\ref{prop:minEeff}.

The cavity mean field equations arose in the context of spin systems in solid state physics~\cite{MPCavity,MPBook}. These equations take into account the feedback dependencies by estimating the reaction of all the other `spins'/variables when a single spin is removed from the system, thereby leaving a `cavity'. This leads to a considerable simplification by utilizing the fact that the system of variables are fully connected. The local susceptibility matrix $\CHI$, a common quantity in physics, measures how stable the solution is to perturbations. This quantity plays a key role in such systems ~\cite{RMSCavity}. In particular, in the asymptotic limit of large $M$ and $N$, certain quantities (e.g. MSE and average local susceptibility, $\overline{\chi}(\x)$) converge, i.e. become independent of the detailed realization of the matrix $\H$. 
In this limit, a sudden increase in susceptibility signals the error prone phase.

These results are equivalent to those obtained by~\cite{Kabashima09,Ganguli10} with the replica approach. These approaches begin with a finite temperature statistical mechanics model. In order to make a connection with these studies, one should replace $\overline{\chi}$ by the quantity $\beta\Delta Q$  where  $\beta$ is a quantity playing the role of inverse temperature. The quantity $\Delta Q$ could be defined as 
\begin{equation}
\Delta Q\equiv[ \langle (u-\langle u\rangle)^2\rangle]^{\mathrm{av}}_{\x_0,\H,\zeta}.
\label{eq:dQ}
\end{equation}
where $\langle \cdots \rangle$ is the average over `thermal' fluctuations of $\u$ with in the ensemble $P_\beta$:
\begin{equation}
P_\beta(\u | \x_0,\H,\zeta) = \frac{1}{Z(\x_0,\H,\zeta)} e^{-\beta\E_(\u;\x_0,\H,\zeta)}.
\label{eq:prob-eff}
\end{equation}
The quantity $\Delta Q$ is nothing but `thermal' fluctuations in $\u$ and $\beta \Delta Q$ can in fact be identified as a local susceptibility due to the fluctuation-dissipation theorem~\cite{Kubo66}. Our results are obtained in the limit $\beta\rightarrow\infty$, where the probability distribution becomes peaked near the minimum, making it into an optimization problem. The local susceptibility, however remains well-defined in this limit.

\section{Ridge Regression via Singular Value Decomposition}\label{app:SVD}
For the sake of completeness, in this appendix, we derive Eqs. \eqref{eq:min-ridge-extreme} and \eqref{eq:sxi-self-consistent-ridge-extreme} in section~\ref{sec:ridge} using a singular value decomposition. Elementary derivation leads us to an explicit expression:
\begin{equation}
\hat{\x}= \frac{\H^\mathrm{T}\H}{\sigma^2}\Big[\frac{\H^\mathrm{T}\H}{\sigma^2}+\lambda\mathbf{I}_N\Big]^{-1}\x_0=\sum_{i=1}^M\frac{s^2_i}{s^2_i+\lambda\sigma^2}\mathbf{\cal V}_i(\mathbf{\cal V}^\mathrm{T}_i\x_0).
\label{eq:ridge-solution}
\end{equation}
where we use the singular vector basis of the matrix $\H$, with $\s_i$ being the non-zero singular values, and $\mathbf{\cal V}_i$ the corresponding right singular vectors. When we take the limit of vanishing $\sigma^2$, we just have a projection of the $N$ dimensional vector $\x_0$ to an $M$-dimensional projection spanned by $\mathbf{\cal V}_i$'s.
In other words
\begin{equation}
x_a= \sum_{b=1}^N\sum_{i=1}^M{\cal V}_{ia}{\cal V}_{ib}x_{0a}=\sum_{a=1}^NP_{ab} x_{0a}
\label{eq:ridge-solution-lim}
\end{equation}
$\mathbf{P}$ being the projection matrix.
For random $\H$, $\mathbf{\cal V}_i$'s are just a random choice of $M$ orthonormal vectors. Thus, the properties of the estimate depends on the statistics of the projection matrix to a random $M$-dimensional subspace.
\begin{equation}
[P_{ab}]^{\mathrm{av}}_{\H}= \sum_{i=1}^M[{\cal V}_{ia}{\cal V}_{ib}]^{\mathrm{av}}_{\H}=\sum_{i=1}^M\frac{\delta_{ab}}{N}=\alpha\delta_{ab}
\implies[\hat x_a]^{\mathrm{av}}_{\H}=\alpha x_{0a}
\label{eq:ridge-solution-av}
\end{equation}
For variance, we need to think of second order moments of the matrix elements of $\mathbf{P}$, particularly, $[P_{ab}P_{ac}]^{\mathrm{av}}_{\H}$. We could parametrize $[P_{ab}P_{ac}]^{\mathrm{av}}_{\H}=A\delta_{bc}+B\delta_{ab}\delta_{bc}$. Since 
$\mathbf{P}$ is a projection operator, $\mathbf{P}^2=\mathbf{P}$ and it is a symmetric matrix. Hence,
\begin{equation}
 \sum_a[P_{ab}P_{ac}]^{\mathrm{av}}_{\H}=\sum_a[P_{ba}P_{ac}]^{\mathrm{av}}_{\H}=[P_{bc}]^{\mathrm{av}}_{\H}=\alpha\delta_{bc}.
 \label{eq:proj-constr}
 \end{equation}
 In the limit of $M,N\rightarrow0$ with $\alpha$ fixed, the distribution of $P_{aa}$ gets highly concentrated around the mean $\alpha$. As a result,
 \begin{equation}
 [P_{aa}P_{aa}]^{\mathrm{av}}_{\H}\approx(\sum_a[P_{aa}]^{\mathrm{av}}_{\H})^2=\alpha^2.
 \label{eq:paa-largen-constr}
 \end{equation}
Using the two constraints, represented by Eqs.~\eqref{eq:proj-constr} and~\eqref{eq:paa-largen-constr}, we can determine $A$ and $B$, in the large $M,N$ limit, leading to,
\begin{equation}
 [P_{ab}P_{ac}]^{\mathrm{av}}_{\H}\approx \frac{\alpha(1-\alpha)}{N}\delta_{bc}+\alpha^2\delta_{ab}\delta_{bc}.
 \label{eq:proj-covar}
 \end{equation}
The variance is now given by, 
\begin{align}
[&\hat x_{0a}^2]^{\mathrm{av}}_{\H}-([\hat x_{0a}]^{\mathrm{av}}_{\H})^2\nonumber\\
&= \sum_a[ \frac{\alpha(1-\alpha)}{N}\delta_{bc}+\alpha^2\delta_{ab}\delta_{bc}]x_{0b}x_{0c} -(\alpha x_{0a})^2\nonumber\\
 &=(1-\alpha)\alpha\rho\big[x_0^2\big]^{\mathrm{av}}_{x_0}
\label{eq:ridge-solution-var}
\end{align}
recovering our earlier result.

\end{appendices}

\bibliographystyle{spmpsci}      % mathematics and physical sciences
\bibliography{ENetJStatPhys,ENetExtra}   % name your BibTeX data base

\end{document}